\documentclass[twocolumn]{aastex63}
\usepackage{lineno}
\usepackage[caption=false]{subfig}
\usepackage{multirow}
\usepackage{graphicx}
\usepackage{booktabs}
\usepackage{threeparttable}
\usepackage{amsmath,amssymb}
\usepackage[T1]{fontenc}
\usepackage{CJKutf8}
\usepackage{rotating}

\graphicspath{{./}{figures/}}

\newcommand{\xmm}{{\it XMM-Newton}}
\newcommand{\chandra}{{\it Chandra}}
\newcommand{\XRT}{{\it Swift-XRT}}

\begin{document}
\begin{CJK}{UTF8}{gbsn}

\title{X-ray Emission Properties of a Compact Symmetric Object Sample}

\correspondingauthor{Jin Zhang}
\email{j.zhang@bit.edu.cn}

\author[0000-0002-4789-7703]{Ying-Ying Gan}
\affiliation{School of Physics, Beijing Institute of Technology, Beijing 100081, People's Republic of China; j.zhang@bit.edu.cn}

\author[0000-0002-9728-1552]{Su Yao}
\affiliation{National Astronomical Observatories, Chinese Academy of Science, Beijing 100101, China}

\author[0009-0006-4551-8235]{Tan-Zheng Wu}
\affiliation{School of Physics, Beijing Institute of Technology, Beijing 100081, People's Republic of China; j.zhang@bit.edu.cn}

\author[0000-0001-6863-5369]{Hai-Ming Zhang}
\affiliation{Guangxi Key Laboratory for Relativistic Astrophysics, School of Physical Science and Technology, Guangxi University, Nanning 530004, People's Republic of China}

\author[0000-0003-3554-2996]{Jin Zhang\dag}
\affiliation{School of Physics, Beijing Institute of Technology, Beijing 100081, People's Republic of China; j.zhang@bit.edu.cn}

\begin{abstract}

We present a comprehensive analysis of the X-ray observations obtained from \xmm\, and \chandra\, for a sample of bona-fide Compact Symmetric Objects (CSOs) to investigate their X-ray emission properties. Ultimately, we obtain 32 effective X-ray observational spectra from 17 CSOs. Most spectra can be well described by an absorbed single power-law model, with the exception of 6 spectra requiring an additional component in the soft X-ray band and 2 spectra exhibiting an iron emission line component. The data analysis results unveil the diverse characteristics of X-ray emission from CSOs. The sample covers X-ray luminosity ranging within $10^{40}-10^{45}$ erg s$^{-1}$, intrinsic absorbing column density ($N_{\rm H}^{\rm int}$) ranging within $10^{20}-10^{23}$ cm$^{-2}$, and photon spectral index ($\Gamma_{\rm X}$) ranging within 0.75--3.0. None of the CSOs in our sample have $N_{\rm H}^{\rm int}$ > $10^{23}\rm~cm^{-2}$, indicating that the X-ray emission in these CSOs is not highly obscured. The distribution of $\Gamma_{\rm X}$ for these CSOs closely resembles that observed in a sample of radio-loud quasars and low-excitation radio galaxies (RGs). In the radio--X-ray luminosity panel, these CSOs exhibit a distribution more akin to FR I RGs than FR II RGs, characterized by higher luminosities. The positive correlation between $\Gamma_{\rm X}$ and the Eddington ratio, which has been noted in radio-quiet active galactic nuclei, is not observed in these CSOs. These findings suggest that although the contribution of the disk-corona system cannot be completely ruled out, jet/lobe radiation likely plays a dominant role in the X-ray emission of these CSOs.

\end{abstract}


\section{Introduction}

Compact symmetric objects (CSOs) are a special type of radio-loud active galactic nuclei (RL-AGNs), characterized by their symmetrical radio morphology with a compact projection size of $\leq$1 kpc (\citealp{1980ApJ...236...89P, 1994ApJ...432L..87W, 1996ApJ...460..612R}). The symmetrical radio morphology exhibited by CSOs resembles that of classical radio galaxies (RGs), thus earning them miniature versions of RGs. It is widely accepted that the jets in CSOs have a large viewing angle and are minimally influenced by relativistic beaming effects, thereby accounting for their symmetrical radio morphology (\citealp{1980ApJ...236...89P, 1994ApJ...432L..87W, 1996ApJ...460..612R}). Based on the perfectly compact and symmetrical radio morphology, certain AGNs are classified as CSOs; however, these CSOs also exhibit properties of relativistic jets, such as PKS 1543+005, CTD 135, and PKS 1413+135 (\citealp{2021ApJ...907...61R, 2021RAA....21..201G, 2022ApJ...939...78G, 2024ApJ...961..240K}). Recently, \cite{2024ApJ...961..240K} introduced two additional criteria to redefine CSOs: low variability and low apparent speeds along the jets. Based on these criteria, they compiled a comprehensive catalog of ``bona fide'' CSOs. 

The reason for the compactness of CSOs is still debated. Both the derived age based on the advance speed of lobe/hotspot and the synchrotron-loss timescale indicate that CSOs are young, less than $10^4$ yr (\citealp{1996ApJ...460..612R, 1998A&A...337...69O, 2000ApJ...541..112T, 2003PASA...20...19M, 2012ApJS..198....5A}). Therefore, it is believed that CSOs represent AGNs in their early stages of evolution and will eventually evolve into classical RGs (\citealp{1980ApJ...236...89P, 1995A&A...302..317F, 1996ApJ...460..634R}). Alternatively, the compactness of CSOs may be attributed to their existence in dense environments (e.g., \citealp{1984AJ.....89....5V, 1997ASPC..121..672B, 2017ApJ...836..174C}).

The physical origin of X-ray emission in CSOs is also a subject of debate. In the radio--X-ray flux plane, high-excitation radio galaxies (HERGs) and low-excitation radio galaxies (LERGs) exhibit distinct distributions. Similarly, compact radio sources also display this distribution pattern (Figure 3 in \citealp{2014MNRAS.437.3063K}). HERGs are characterized by radiatively efficient accretion disks with an Eddington ratio ($R_{\rm Edd}$) > 0.01, while LERGs possess radiatively inefficient disks with $R_{\rm Edd}$ < 0.01 (\citealp{2014ARA&A..52..589H}). This indicates two distinct modes of X-ray emission may be present in compact radio sources: jets/lobes and the disc-corona system. The X-ray emission from the disc-corona system is generally obscured by the torus and often accompanied by iron emission line (e.g., \citealp{2007MNRAS.382..194N, 2012A&A...537A..86I, 2014MNRAS.441.3622R}). However, in the case of CSOs with dimensions comparable to or smaller than the torus, X-ray emission from jets/lobes may also be obscured by the torus, leading to the generation of iron emission line (\citealp{2024ApJ...966..201K}). In this scenario, X-rays are produced via the inverse Compton (IC) process involving relativistic electrons in their nucleus-jet or mini-lobes; furthermore, distinguishing between these two IC processes remains challenging. \cite{2008ApJ...680..911S} proposed a simple dynamical model to describe the evolution of a relativistic jet-cocoon system, which was successfully applied by \cite{2010ApJ...715.1071O} to fit the broadband spectral energy distributions (SEDs) of gigahertz-peaked spectrum objects (GPSs). According to this model, X-rays are produced through the IC scattering of relativistic electrons in lobes. Furthermore, this model forecasts the observable $\gamma$-ray emission from some compact radio sources. Now three bona fide CSOs (PKS 1718--649, NGC 3894, TXS 0128+554) defined in \cite{2024ApJ...961..240K} have been detected in high-energy $\gamma$-ray band (\citealp{2016ApJ...821L..31M, 2020A&A...635A.185P, 2020ApJ...899..141L, 2024RAA....24b5018G}). Additionally, one CSO, NGC 4278, has been observed in the very-high-energy (VHE) $\gamma$-ray band (\citealp{2024ApJS..271...25C}). Alternatively, X-rays could originate from the thermal emission of the large-scale environment (\citealp{1998ApJ...501..126H, 2000AJ....119..478O}). All these thermal and non-thermal radiation processes could contribute to the total X-ray emission observed in CSOs, making it challenging to disentangle these contributions. 

In this paper, we carefully select a sample of authentic CSOs from \cite{2024ApJ...961..240K} and conduct a comprehensive analysis of the \chandra\, and \xmm\, observations for these CSOs, while also integrating multi-frequency radio data to explore their X-ray emission properties. The detailed description of the selected sample is presented in Section 2. Section 3 and Section 4 present the data analysis of X-ray observations along with the corresponding results. Discussion on the obscuration and origin of X-ray emission in these CSOs is given in Section 5. A summary is provided in Section 6. Throughout, the cosmological parameters $H_0=71$ km~s$^{-1}$~Mpc$^{-1}$, $\Omega_{\rm m}=0.27$, and $\Omega_{\Lambda}=0.73$ are adopted.

\section{Sample Selection}

Based on the redefinition of CSOs, \cite{2024ApJ...961..240K} have compiled a catalog of bona fide CSOs, including 79 CSOs. Among these CSOs, 27 CSOs have been observed by \chandra\, and/or \xmm. It should be noted that NGC 4278 is also included in this bona fide CSO catalog and has the \chandra\, and \xmm\, observations. However, it has been reported to be associated with a VHE $\gamma$-ray source 1LHAASO J1219+2915, and the integral flux during active periods is $\sim 7$ times higher than that during quiet periods (\citealp{2024ApJS..271...25C, 2024ApJ...971L..45C}). Additionally, \XRT\, observations during the active phase in the TeV band reveal an X-ray flux more than one order of magnitude higher than that detected by \chandra\, approximately 12 yr earlier (\citealp{2024arXiv240500347L}). Furthermore, its flux density at 6 cm from 1972 to 2003 shows significant variability on timescales of years (\citealp{2005ApJ...622..178G}). So NGC 4278 is excluded from our sample. CSOs OQ 208, JVAS J1511+0518, S4 2021+61, and NGC 7674 have been reported as X-ray obscured AGNs and extensively studied alongside hard-X-ray observation data (\citealp{2017MNRAS.467.4606G, 2019ApJ...884..166S, 2023ApJ...948...81S}). In this paper, we will not reanalyze their observation data. However, the findings reported for these four CSOs are taken into consideration in the discussion presented in Section \ref{discu}. More detailed information about these four obscured CSOs can be found in Appendix \ref{append_4CSO}. Our main focus in this study is on the remaining 22 CSOs presented in Table \ref{tab-22CSO}.

The information of the 22 CSOs, together with the four obscured CSOs, is given in Table \ref{tab-22CSO}, including their redshift, linear size, angular size, and Eddington ratio. And, the information is taken from \cite{2002ApJ...579..530W}, \cite{2011A&A...528A.110F}, \cite{2020MNRAS.491...92L}, \cite{2020ApJ...892..116W}, \cite{2021ApJ...922...84B}, and \cite{2024ApJ...961..240K}.
The highest redshift among these CSOs is 0.66847 for CSO 0108+388, while half of the sources have a redshift below 0.15. All sources have a linear size smaller than 750 pc, and half of them possess a linear size of less than 150 pc. The majority of these sources exhibit an angular size below 100 milliarcseconds (mas), except for one source with an angular size exceeding 1 arcsec. This small angular extent implies that even the X-ray satellites like \chandra\, are unable to resolve the core and jet structures within these CSOs. Therefore, all CSOs are treated as point-like sources during data analysis.

\section{X-ray Observations and Data Analysis}

Among the 22 CSOs, \chandra\, has observed 13 CSOs (Table \ref{tab-chandra}), while \xmm\, has conducted observations on 16 CSOs (Table \ref{tab-XMM}). There is an overlap of observations between \chandra\, and \xmm\, for 7 CSOs. Additionally, several CSOs have been observed more than three times. All source positions are obtained from \cite{2024ApJ...961..240K}. 

\subsection{X-ray Data Reduction}

This work employs a list of \chandra\, datasets, obtained by the \chandra\, X-ray Observatory, contained in~\dataset[doi:10.25574/cdc.302]{https://doi.org/10.25574/cdc.302}. The \chandra\, Advanced CCD Imaging Spectrometer S-array (ACIS-S) is used for all \chandra\, observations, with the exception of B3 0402+379, which employs the I-array (ACIS-I). We process \chandra\, observation data using CIAO (version 4.15) with calibration database CALDB (version 4.10.4), and generate the level-2 event file following standard procedures. The source spectrum is extracted from a circle with a radius of $\sim7^{\prime\prime}$ centered on the source position, while the background spectrum is extracted from an annular region with inner and outer radii of $10^{\prime\prime}$ and $20^{\prime\prime}$ respectively, also centered on the source position. The total counts and net counts within the range of 0.5 to 7.5 keV in the source extraction region for each observation are provided in Table \ref{tab-chandra}. These values are obtained using XSPEC (version 12.13.0c).

For the \xmm\, observations, we process the observation data using the \xmm\, SCIENCE ANALYSIS SYSTEM (SAS, version -1.3) and the latest calibration files following standard procedures. \xmm\, carries three X-ray CCD cameras, including the European Photon Imaging Camera (EPIC)-PN and two EPIC-MOS (MOS1 and MOS2) CCD arrays. Spectra are generated using observation data from both the PN and MOS CCD arrays. Time intervals with background count rate exceeding 0.4 counts s$^{-1}$ for PN and 0.35 counts s$^{-1}$ for MOS are excluded from analysis. The resulting clean exposure times for each observation from the PN and MOS instruments are summarized in Table \ref{tab-XMM}. Source photons are extracted from a circle with a radius of $\sim32^{\prime\prime}$ centered on the source position, while background photons are extracted from a similar circle within a nearby no-source region on the same CCD. As both PKS 1117+146 and PKS 1934--63 have an object nearby, the source and background regions with a radius of $\sim16^{\prime\prime}$ are used for the two CSOs. The total counts and net counts within the range of 0.5 to 7.5 keV in the source extraction region for each observation from PN and MOS are presented in Table \ref{tab-XMM}. These values are obtained using XSPEC (version 12.13.0c). MOS1 and MOS2 are identical instruments with similar instrumental response functions. We first combine the spectra of MOS1 and MOS2 into one spectrum for each source, and then fit the PN spectrum and the combined MOS spectrum simultaneously.

The pileup effect has been examined for all observations. For each observation conducted by \chandra\,, a pileup map is generated using the CIAO task $pileup\_map$\footnote{http://cxc.harvard.edu/ciao/ahelp/acis\_pileup.html}. In the produced pileup map, any pixel with a value exceeding 0.2 counts per frame indicates the presence of pileup. Regarding \xmm\, observations, a bright point source with a count rate surpassing 0.7 counts s$^{-1}$ should be assessed for the occurrence of pileup using the $epatplot$ task. Two \chandra\, observations suffer from the impact of pileup, namely Obs-ID 3055 for JVAS J1234+4753 and Obs-ID 16623 for PKS 1718--649, while no consideration needs to be given to the occurrence of pileup in \xmm\, observations.

During the data analysis of Obs-ID 3055 for JVAS J1234+4753 and Obs-ID 16623 for PKS 1718--649, the source spectrum is extracted from an annular region based on the generated pileup map. This extraction process excludes any pixel regions that have a count value surpassing 0.2 counts per frame. In Obs-ID 3055 for JVAS J1234+4753, where 9 pixels exceed this threshold, an annular region with inner and outer radii of approximately $1^{\prime\prime}$ and $7^{\prime\prime}$ respectively is employed for extracting the source spectrum. Similarly, in Obs-ID 16623 for PKS 1718--649, where 4 pixels surpass the count limit in the pileup map, an annular region with inner and outer radii of approximately $0.7^{\prime\prime}$ and $7^{\prime\prime}$ respectively is utilized to extract the source spectrum. The background spectra are obtained from an annular region with inner and outer radii of $10^{\prime\prime}$ and $20^{\prime\prime}$. The total and net counts within the range of 0.5 to 7.5 keV in the source extracted regions for both observations are presented in Table 4.

The factors, such as low X-ray signals observed from the source, short exposure time during observation, or persistently high background activities ultimately hinder our ability to extract net counts or generate an effective energy spectrum for certain observations. For specific details regarding these issues in each observation and individual source, please refer to Appendix \ref{append_indiv}, as well as Tables \ref{tab-chandra} and \ref{tab-XMM}.

\subsection{Spectral Analysis}

We perform X-ray spectral analysis using XSPEC (version 12.13.0c). For observations with a sufficient number of net counts, we utilize the $grppha$ tool to group the source spectrum, ensuring a minimum of 20 counts per energy bin and apply the $\chi^2$ statistic for spectral analysis. In cases where the net counts are insufficient, we set a minimum of 5 counts per energy bin to group the source spectrum and employ the C-statistic for spectral analysis. For S4 1943+54, which exhibits extremely low counts, we reduce the minimum number of counts per energy bin to 2 and continue to use the C-statistic. It is important to note that the C-statistic is not valid when background counts are non-negligible. Therefore, despite the low net counts observed for PKS 1117+14, we set a minimum of 15 counts per energy bin and maintain the use of the $\chi^2$ statistic for its spectral analysis. 

We first adopt a simple absorbed power-law model to fit the X-ray spectrum of all sources in our sample, considering both Galactic and host galaxy absorption, i.e.,
\begin{equation}
N(E) = AE^{-\Gamma_{\rm X} \exp \left\{-N_{\rm H}^{\rm gal} \sigma(E) - N_{\rm H}^{\rm int} \sigma \left[E(1+z)\right] \right\}},
\end{equation}
where $A$ is the normalization at 1 keV, $\Gamma_{\rm X}$ is the photon spectral index, $\sigma$ is the photo-electric cross section, and $N_{\rm H}^{\rm gal}$ and $N_{\rm H}^{\rm int}$ are the absorbing column densities of Galactic and host galaxy, respectively. In XSPEC, the model can be expressed as \texttt{tbabs$*$ztbabs(powerlaw)}. The value of $N_{\rm H}^{\rm gal}$ is fixed at the reported value in \cite{2016A&A...594A.116H}, while $N_{\rm H}^{\rm int}$ is kept free during the spectral fitting. 

For sources with multiple observations available, we initially conduct spectral fitting on the combined spectrum of these observations to establish a constraint for $N_{\rm H}^{\rm int}$. Subsequently, we fix the value of $N_{\rm H}^{\rm int}$ while allowing $\Gamma_{\rm X}$ and $A$ to vary freely during the spectral fitting of each individual observation, as exemplified by TXS 0128+554, B3 0402+379, B3 0710+439, S5 1946+70, and PKS 1718--649. However, for certain sources where the combined spectrum fails to provide reliable parameter constraints, we opt to fit their individual spectra directly, keeping $N_{\rm H}^{\rm int}$, $\Gamma_{\rm X}$, and $A$ as free parameters, such as NGC 3894, JVAS J1234+4753, CTD 93, and PKS 1934--63. All spectral fittings are performed within the 0.5--7.5 keV energy band. Further details regarding each source can be found in Tables \ref{Joint-fit}, \ref{tab-Result}, and in Appendix \ref{append_indiv}.

Significant residuals are observed in the soft X-ray band for the spectra of PKS 1718--649\footnote{The three \chandra\ observations of PKS 1718-649 did not reveal significant residuals in the soft X-ray band, likely attributable to the limited size of the extracted regions.}, NGC 3894, and PKS B1345+125 when fitting these spectra with the absorbed power-law model. This suggests the presence of an additional spectral component. Extended X-ray emission on the kpc scale, beyond the size of the CSOs, has been previously detected in these sources (\citealp{2008ApJ...684..811S, 2018A&A...612L...4B, 2021ApJ...922...84B}). We thus consider a thermal plasma component (\texttt{APEC} in XSPEC) to optimize the spectral fitting. In addition, 2 spectra exhibit an emission line feature at around 6.4 keV, necessitating the inclusion of a Gaussian model (\texttt{zgauss} in XSPEC) for spectral fitting. For specific information regarding individual sources, please refer to the Appendix \ref{append_indiv}.

\section{Results of X-ray Data Analysis}

By conducting a comprehensive analysis of the observational data, we have successfully derived 18 valid X-ray spectra from 12 \chandra\, observational CSOs and obtained 14 valid X-ray spectra from 12 \xmm\, observational CSOs. Among these sources, 7 CSOs are included in both \chandra\, and \xmm\, observations; therefore, we obtain at least one reliable observational spectrum for each of the 17 CSOs through our meticulous data analysis. The majority of the spectra can be adequately fitted using a simple absorbed power-law model, except for 6 spectra from 3 CSOs showing residuals in the soft X-ray band and requiring the second component, and 2 spectra from 2 CSOs exhibiting an iron emission line component. The Appendix \ref{append_indiv} and Tables \ref{Joint-fit}, \ref{tab-Result} provide additional details for individual CSOs. For sources with multiple observations, the values of $N_{\rm H}^{\rm int}$, $\Gamma_{\rm X}$, and $L_{\rm 2-10~keV}$ (the corrected luminosity in the 2--10 keV band) used in all distribution and correlation analyses are determined based on a selection criterion. Specifically, we first consider the fitting results from joint spectral analysis. If joint spectral fitting is not available, we prioritize the results that best constrain $N_{\rm H}^{\rm int}$, followed by selecting the observation with the maximum net photon counts. The main findings are summarized below.

\begin{figure*}
 \centering
   \includegraphics[angle=0,scale=0.35]{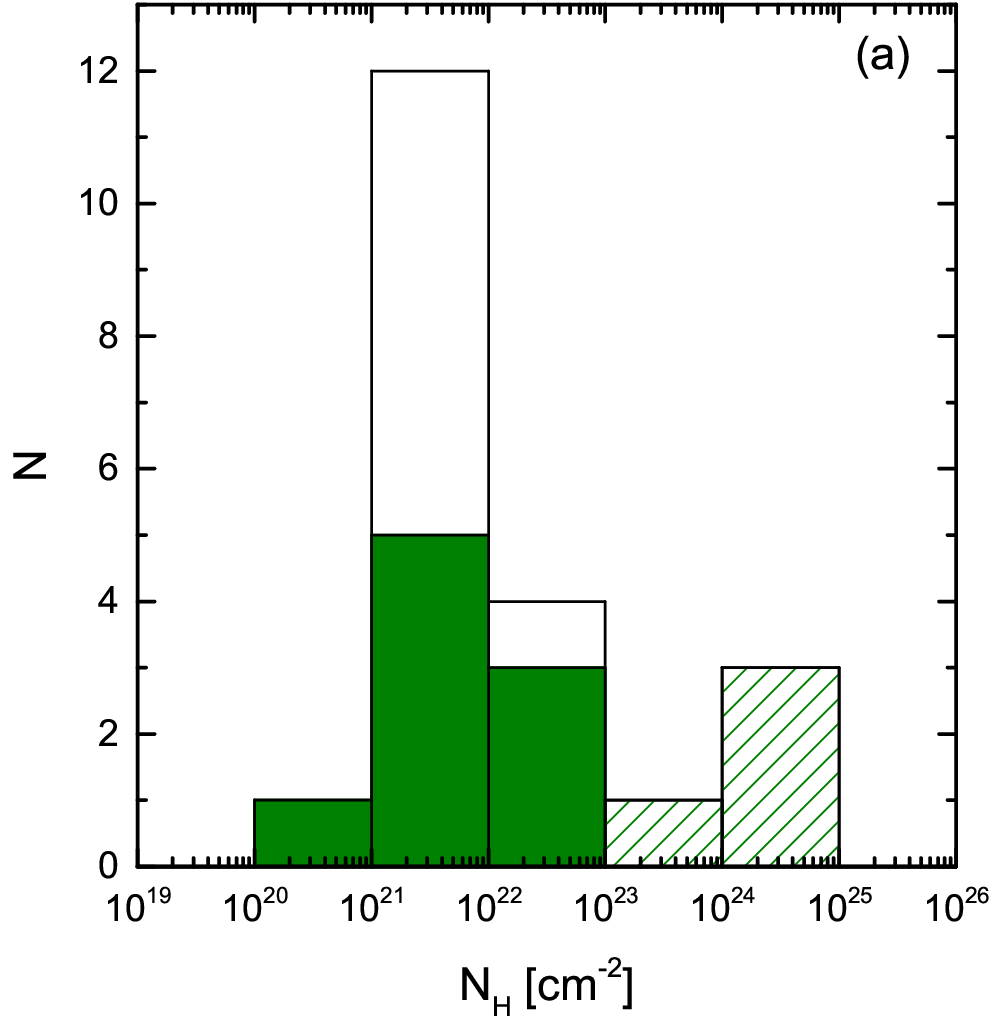}
   \includegraphics[angle=0,scale=0.35]{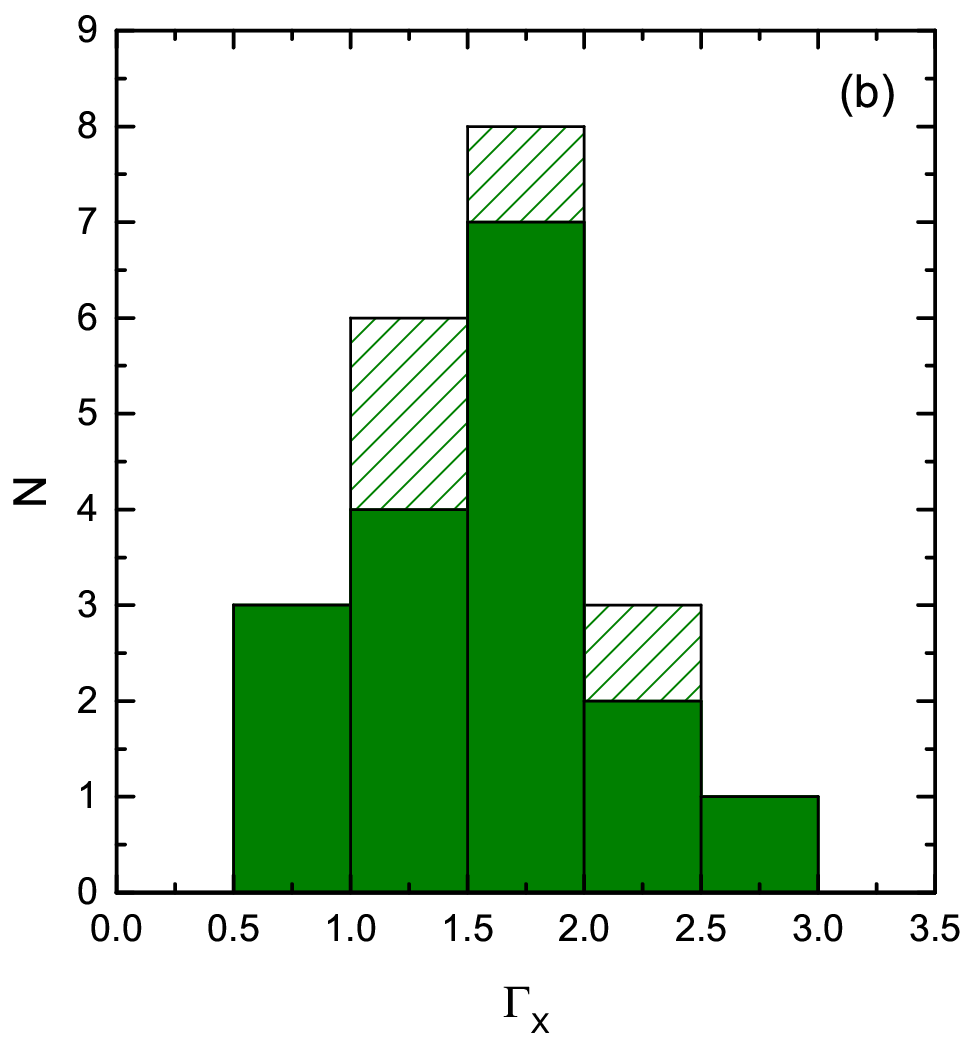}
   \includegraphics[angle=0,scale=0.35]{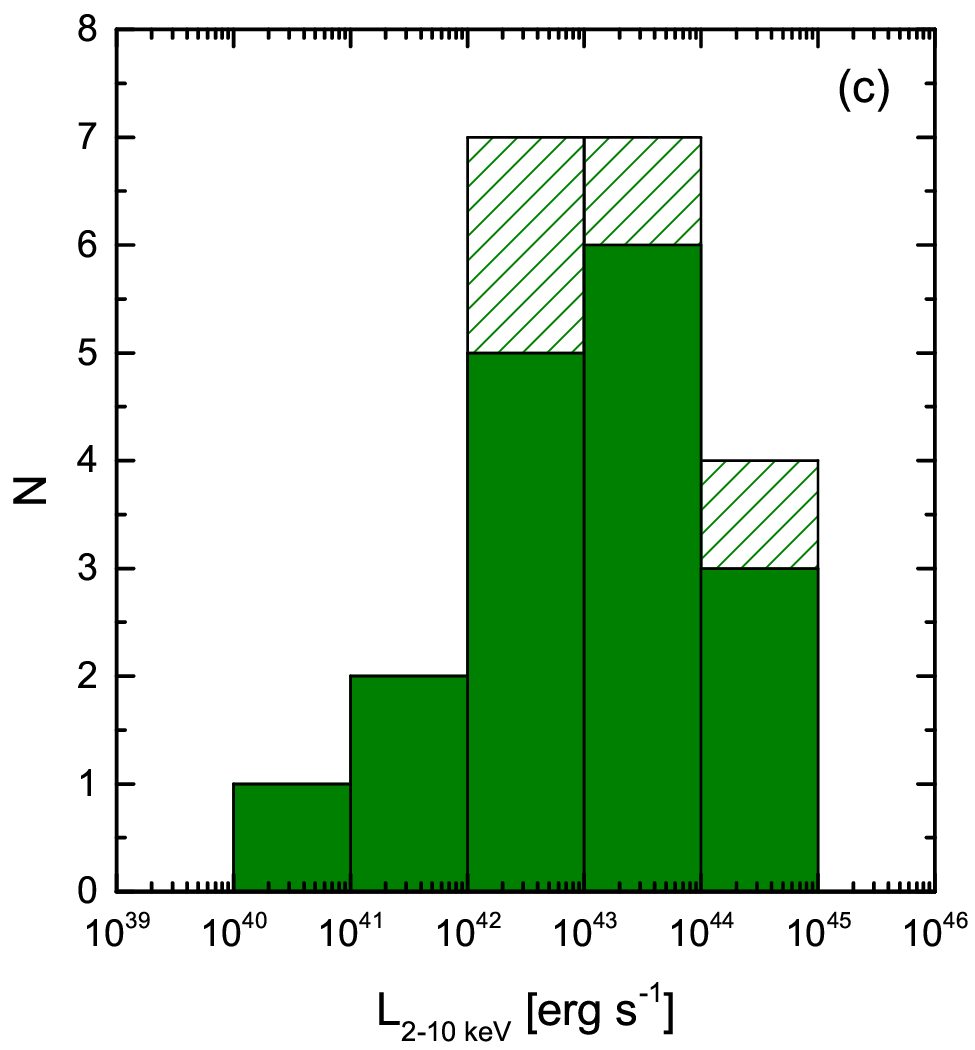}
\caption{The distributions of $N_{\rm H}^{\rm int}$ (\emph{Panel}-(a)), $\Gamma_{\rm X}$ (\emph{Panel}-(b)), and $L_{\rm 2-10~keV}$ (\emph{Panel}-(c)) for the CSO sample, where the diagonal shaded areas indicate the distributions of 4 obscured CSOs. The data of the 4 obscured CSOs are taken from \cite{2017MNRAS.467.4606G} and \cite{2019ApJ...884..166S, 2023ApJ...948...81S}. The empty area in the \emph{Panel}-(a) indicates that only an upper-limit value of $N_{\rm H}^{\rm int}$ is obtained.}
\label{dis_Nh_Gamm_Lx}
\end{figure*}

\begin{figure*}
 \centering
   \includegraphics[angle=0,scale=0.5]{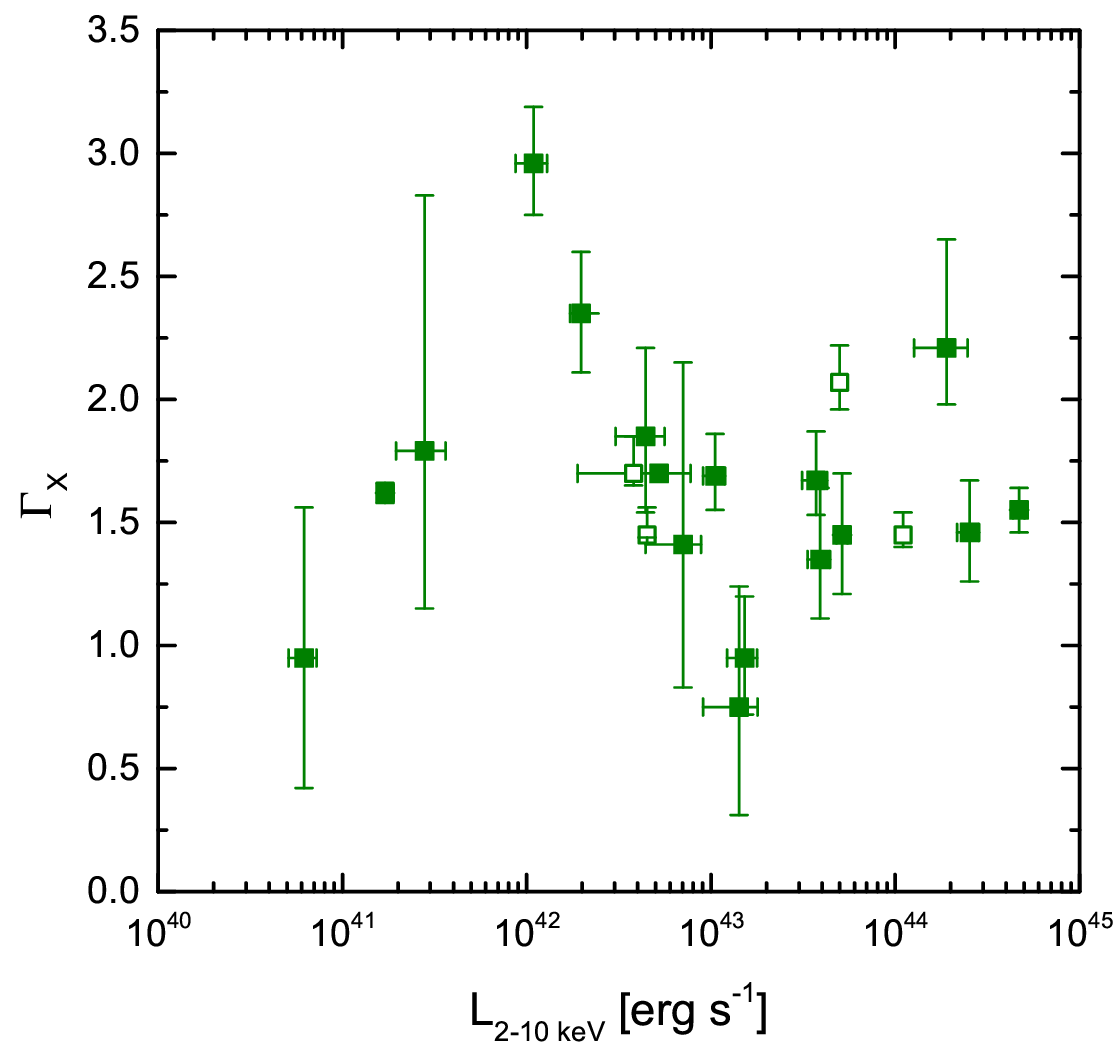}
\caption{$\Gamma_{\rm X}$ as a function of $L_{\rm 2-10~keV}$ for the CSO sample. The solid squares represent the 17 CSOs analyzed in this work while the opened squares represent the 4 obscured CSOs. The data of the 4 obscured CSOs are taken from \cite{2017MNRAS.467.4606G} and \cite{2019ApJ...884..166S, 2023ApJ...948...81S}.}
\label{cor_Gamma-Lx}
\end{figure*}

\begin{itemize}

\item{$N_{\rm H}^{\rm int}$.} When comparing these values with the $N_{\rm H}^{\rm gal}$ values for each CSO, it becomes evident that an additional absorption component besides the Galactic absorption is required to accurately fit these X-ray spectra. The distribution of $N_{\rm H}^{\rm int}$ is primarily concentrated between $10^{21}$ and $10^{22}$ cm$^{-2}$, as illustrated in Figure \ref{dis_Nh_Gamm_Lx}(a). Notably, these values are significantly lower than those observed for the four obscured CSOs, which possess an $N_{\rm H}^{\rm int}$ exceeding $10^{23}$ cm$^{-2}$(\citealp{2017MNRAS.467.4606G, 2019ApJ...884..166S, 2023ApJ...948...81S}). 

\item{$\Gamma_{\rm X}$.} The $\Gamma_{\rm X}$ values of the CSOs in our sample span a wide range, from 0.75 to $\sim 3.0$, with a clustering tendency around $\Gamma_{\rm X}\sim1.5-2.0$, as presented in Figure \ref{dis_Nh_Gamm_Lx}(b). Notably, three CSOs (TXS 0128+554, B3 0402+379, JVAS J1234+4753) exhibit discernibly softer spectra compared to the other CSOs, characterized by values of $\Gamma_{\rm X} > 2.0$. Note that individual sources exhibiting a low signal-to-noise ratio tend to have a low $\Gamma_{\rm X}$ value (< 1.4), which is likely attributable to absorption effects.

\item{$L_{\rm 2-10~keV}$.} We calculate their unabsorbed flux in the 2--10 keV band ($F_{\rm 2-10~keV}$) by extrapolating the power-law spectral component in the 0.5--7.5 keV band. And, the corresponding luminosity in the 2--10 keV band ($L_{\rm 2-10~keV}$) is also calculated. As illustrated in Figure \ref{dis_Nh_Gamm_Lx}(c), the values of $L_{\rm 2-10~keV}$ span five orders of magnitude, from $10^{40}$ erg s$^{-1}$ to $10^{45}$ erg s$^{-1}$, clustering around $10^{42}-10^{44}$ erg s$^{-1}$. 

\item{The $L_{\rm 2-10~keV}$--$\Gamma_{\rm X}$ plane.} We plot $\Gamma_{\rm x}$ against $L_{\rm 2-10~keV}$ for the 17 CSOs and the 4 obscured CSOs in Figure \ref{cor_Gamma-Lx}. No correlation is found between $L_{\rm 2-10~keV}$ and $\Gamma_{\rm x}$.

\item{Variability.} In our sample, 9 CSOs have more than one observation, as shown in Tables \ref{tab-Result}. TXS 0128+554, B3 0402+379, and JVAS J1234+4753 consistently exhibit a steep spectrum. $\Gamma_{\rm X}$ and $F_{\rm 2-10~keV}$ of the three sources remain constant within their errors. Similarly, no significant variation in $\Gamma_{\rm X}$ and $F_{\rm 2-10~keV}$ is observed for B3 0710+439 and PKS 1934--63; however, these two CSOs display a flat spectrum. Both CTD 93 and S5 1946+70 have no significant variability in X-ray flux, however there is a change in the spectral shape from soft to hard. Although the flux of NGC 3894 experienced a slight change, two observations obtained by \chandra\, and \xmm\, detectors separated by a duration of 13 yr. As for PKS 1718--649, there are three \chandra\, and three \xmm\, observations available. Although the flux observed in 2014 is nearly twice as high as that in 2010, 2017, 2018, and 2020, the fluxes recorded on June 20th and June 23rd of that year appear to be identical considering errors. While the 2014 observations of PKS 1718--649 yield comparable flux levels, they exhibit distinct $\Gamma_{\rm X}$ values: $\Gamma_{\rm X}=1.70^{+0.09}_{-0.09}$ and $\Gamma_{\rm X}=2.25^{+0.12}_{-0.12}$, respectively. It should be noted that the pileup effect may have influenced the observation (Obs-ID 16623, conducted on June 23rd), potentially contributing to the steeper spectrum with $\Gamma_{\rm X}=2.25^{+0.12}_{-0.12}$, although this effect was accounted for during data analysis.

\item{Soft X-ray exceeds. Six spectra from three CSOs (NGC 3894, PKS B1345+125, and PKS 1718-649) exhibit significant residuals in the soft X-ray band, necessitating the inclusion of a second component to optimize the spectral fits. Considering the extended X-ray emission observed in these sources (\citealp{2008ApJ...684..811S, 2018A&A...612L...4B, 2021ApJ...922...84B}), a thermal plasma model is introduced to account for this excess soft X-ray emission. The fitting results yield a temperature of $kT \sim 0.8$ keV for the spectra of the three sources. Notably, the obtained $kT$ values for NGC 3894 and PKS 1718--649 are consistent with those reported in \cite{2018A&A...612L...4B}, \cite{2021ApJ...922...84B}, and \cite{2024A&A...684A..65B}.}

\end{itemize}

\section{Discussion}\label{discu}   

\subsection{Comparison of Absorbing Columns between CSOs and RGs}

RGs, classified into centre-brightened (FR I) and edge-brightened (FR II) types (\citealp{1974MNRAS.167P..31F}), are also believed to have large viewing angles (\citealp{1995PASP..107..803U}). And CSOs are considered to be the miniature versions of RGs. FR II RGs tend to be highly obscured, with intrinsic absorbing column density generally exceeding $N_{\rm H}^{\rm int}$>$10^{23}$ cm$^{-2}$ (\citealp{2009MNRAS.396.1929H}). In contrast, a study by  \cite{2004ApJ...617..915D} presented \chandra\, and \xmm\, observations of 25 FR I RGs, revealing that only eight sources exhibit excess absorbing column density over the Galactic absorbing column density, with most values in the range of $10^{20}$ cm$^{-2}$ to $10^{21}$ cm$^{-2}$. We compile the $N_{\rm H}^{\rm int}$ values for a sample of RGs from the literature, including the FR I RG sample from \cite{2004ApJ...617..915D} and \cite{2006ApJ...642...96E}, as well as the FR II RG sample from \cite{2009MNRAS.396.1929H}, and compare these with those of CSOs in our sample. 

\begin{figure*}
 \centering
   \includegraphics[angle=0,scale=0.5]{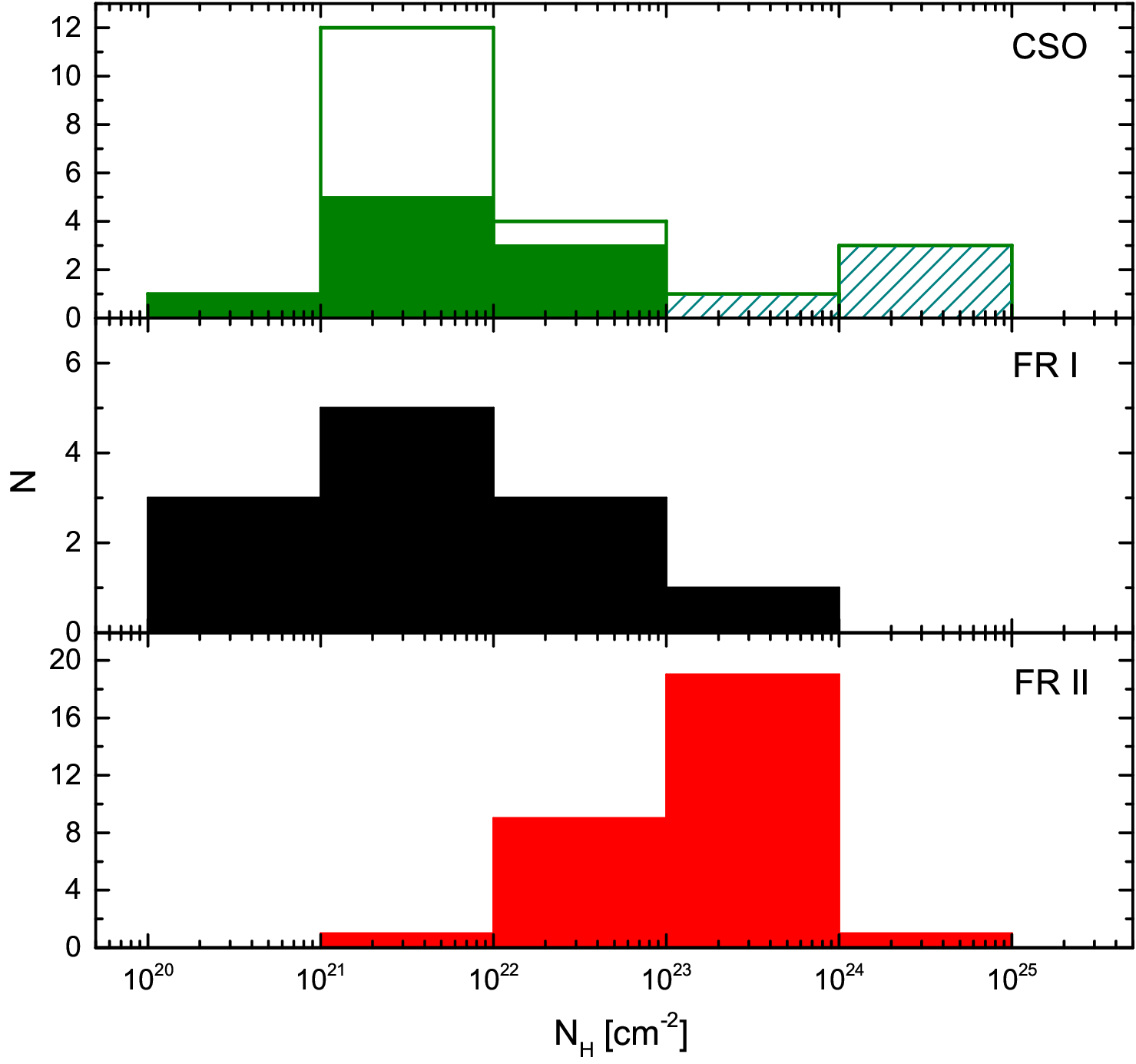}
\caption{Comparison of the $N_{\rm H}^{\rm int}$ distributions. The $N_{\rm H}^{\rm int}$ values of FR I RGs are taken from \cite{2004ApJ...617..915D} and \cite{2006ApJ...642...96E}. The $N_{\rm H}^{\rm int}$ values of FR II RGs are taken from \cite{2009MNRAS.396.1929H}.}
\label{NH_compare}
\end{figure*}

As shown in Figure \ref{NH_compare}, the distribution of $N_{\rm H}^{\rm int}$ for CSOs spans a broad range, encompassing the regions occupied by both FR I and FR II RGs. The $N_{\rm H}^{\rm int}$ values for the majority of CSOs and FR I RGs cluster within the range of $10^{21}$ cm$^{-2}$ to $10^{22}$ cm$^{-2}$, whereas those for most FR II RGs are concentrated between $10^{23}$ cm$^{-2}$ and $10^{24}$ cm$^{-2}$. We further examine the statistical differences between CSOs (excluding upper limit) and FR I RGs as well as FR II RGs using the Kolmogorov--Smirnov test (K--S test), which yields a chance probability of $p_{\rm KS}$. When $p_{\rm KS} > 0.1$, it would strongly suggest no statistical difference between two samples, while it would strongly indicate a statistical difference between two samples if $p_{\rm KS} < 10^{-4}$. We obtain $p_{\rm KS} = 0.15$ between CSOs and FR I RGs and $p_{\rm KS} = 1.75\times10^{-3}$ between CSOs and FR II RGs. So, most CSOs in our sample are not highly obscured; they are more like FR I RGs than FR II RG.

\subsection{Origins of X-ray Emission}

\subsubsection{Comparison of $\Gamma_{\rm X}$ with Other Types of AGNs}

\begin{figure*}
 \centering
   \includegraphics[angle=0,scale=0.5]{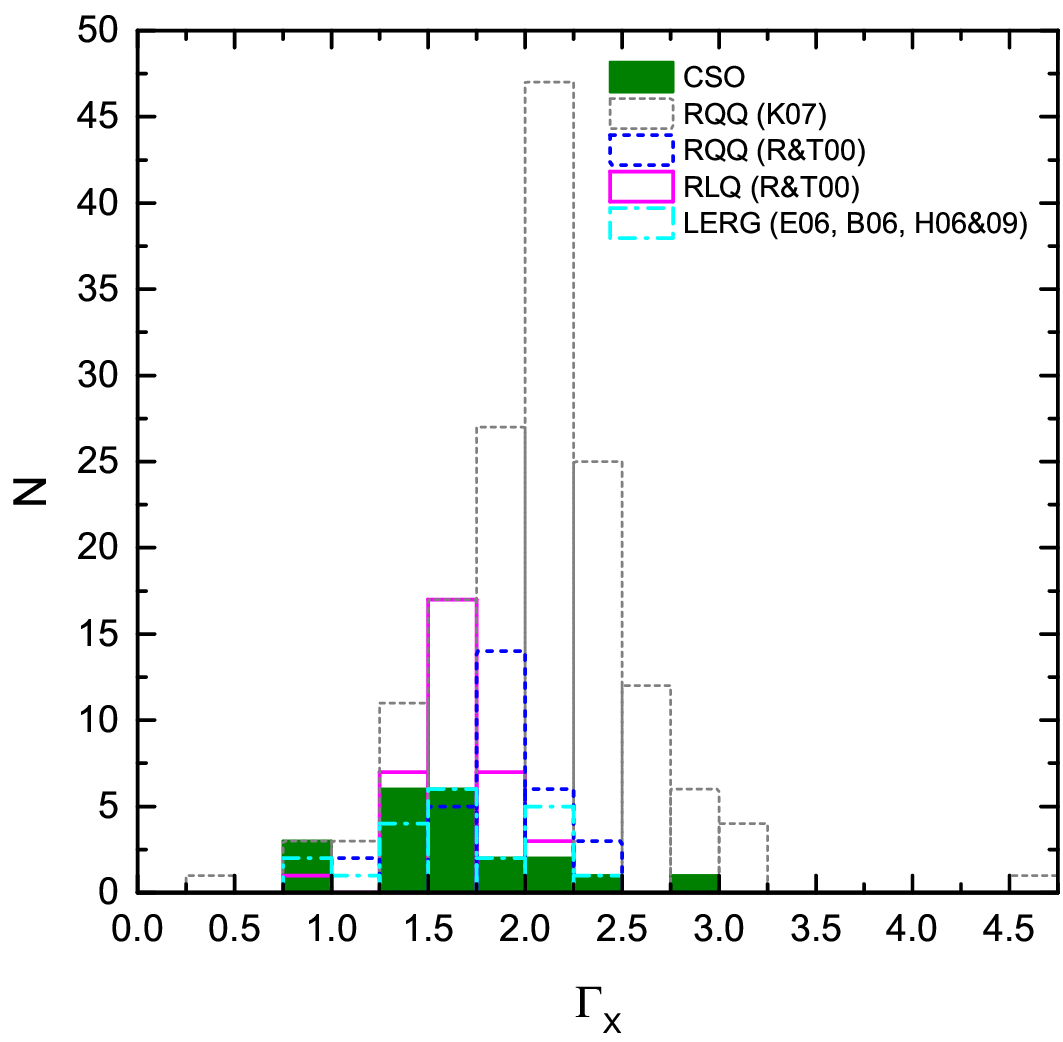}
\caption{The distributions of $\Gamma_{\rm X}$ for various types of AGN samples, where the LERG sample is taken from \cite{2006ApJ...642...96E}, \cite{2006MNRAS.366..339B}, and \cite{2006MNRAS.370.1893H, 2009MNRAS.396.1929H}. The RLQ sample is taken from \cite{2000MNRAS.316..234R}, and the two RQQ samples are taken from \cite{2000MNRAS.316..234R} and \cite{2007ApJ...657..116K}, respectively.}
\label{dis_Gamm}
\end{figure*}

The distribution of the photon spectral index $\Gamma_{\rm X}$ can offer insights into the origin of X-ray emission in CSOs. Previous studies have suggested that different origins of X-rays may correspond to distinct ranges of spectral indices (\citealp{2008ApJ...684..811S, 2009A&A...501...89T}). Therefore, we compile a sample comprising RL quasars (RLQs; \citealp{2000MNRAS.316..234R}) and RQ quasars (RQQs; \citealp{2000MNRAS.316..234R, 2007ApJ...657..116K}), as well as LERGs (\citealp{2006MNRAS.366..339B, 2006ApJ...642...96E, 2006MNRAS.370.1893H, 2009MNRAS.396.1929H}). We compare their distributions of $\Gamma_{\rm X}$ with that of CSOs in our sample, as illustrated in Figure \ref{dis_Gamm}. The average values for $\Gamma_{\rm X}$ are found to be $\Gamma_{\rm X}=1.66\pm0.04$ for RLQs in \cite{2000MNRAS.316..234R}, and $\Gamma_{\rm X}=1.89\pm0.05$ and $\Gamma_{\rm X}=2.03\pm0.31$ for RQQs in \cite{2000MNRAS.316..234R} and in \cite{2007ApJ...657..116K} respectively, while it is $\Gamma_{\rm X}=1.60\pm0.02$ for CSOs in our sample.

As illustrated in Figure \ref{dis_Gamm}, the $\Gamma_{\rm X}$ distribution of CSOs closely resembles that of RLQs and LERGs. We further employ the K-S test to investigate the statistical differences in the $\Gamma_{\rm X}$ distribution between CSOs and RQQs, RLQs, and LERGs. The K--S tests yield $p_{\rm KS} = 0.80$ between CSOs and LERGs, $p_{\rm KS} = 0.30$ between CSOs and RLQs, and $p_{\rm KS} = 6.31 \times 10^{-4}$ between CSOs and RQQs (sample from \citealt{2000MNRAS.316..234R}), while $p_{\rm KS} = 4.51 \times 10^{-5}$ between CSOs and a larger RQQ sample reported in \cite{2007ApJ...657..116K}. Hence, the $\Gamma_{\rm X}$ distribution of CSOs in our sample is more similar to that of RLQs and LERGs, suggesting that the X-ray emission of these CSOs may originate from non-thermal processes and be dominated by nucleus-jet or mini-lobe emission, akin to those observed in RLQs and LERGs.

\subsubsection{Correlations between Radio and X-ray Emission}

Although LERGs and HERGs do not have a one-to-one correspondence with the FR I–FR II categories, most LERGs show FR I RG morphology (\citealp{1979MNRAS.188..111H, 1994ASPC...54..201L, 2009MNRAS.396.1929H}). In FR II RGs, the X-ray emission primarily originates from the disk-corona system, whereas in FR I RGs, it arises from non-thermal processes associated with the jets (\citealp{2000MNRAS.314..359H, 2006A&A...451...35B, 2009MNRAS.396.1929H}). On the radio--X-ray luminosity plane, FR I RGs/LERGs and FR II RGs/HERGs occupy distinct regions, supporting the different dominant origins of their X-ray emission (\citealt{2009A&A...501...89T}, \citealt{2014MNRAS.437.3063K}). To investigate the dominant origin of X-rays in our CSO sample, we compile a sample of FR I and FR II RGs and compare them with our CSO sample in the radio--X-ray luminosity panel. The data of FR I and FR II RGs are taken from \cite{2009A&A...501...89T} and references therein, excluding the upper limit data points. For CSOs, the data at 5 GHz and 8 GHz are taken from the Radio Fundamental Catalog\footnote{http://astrogeo.org/rfc/} (RFC).

\begin{figure*}
 \centering
   \includegraphics[angle=0,scale=0.38]{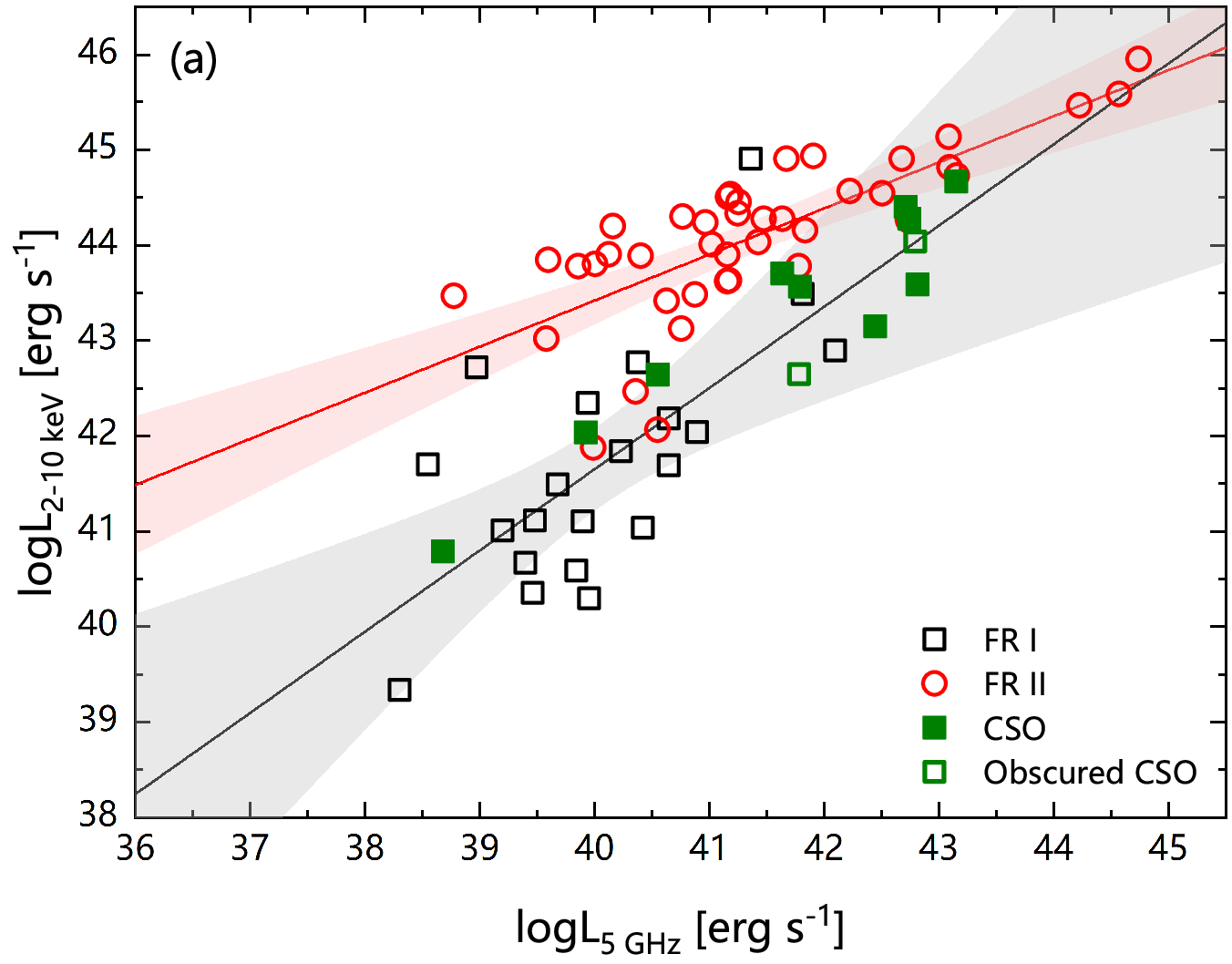}
   \includegraphics[angle=0,scale=0.38]{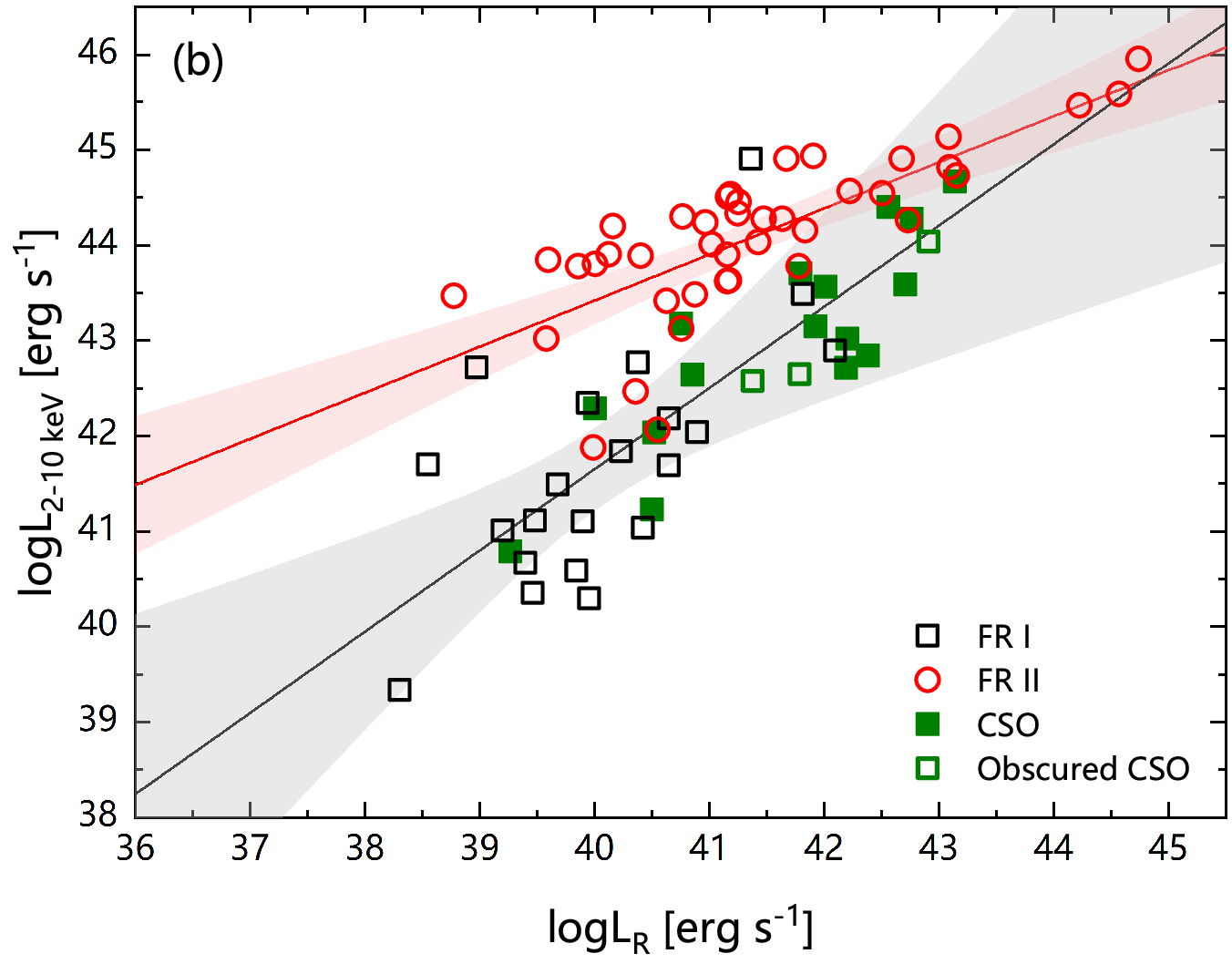}
\caption{\emph{Panel}-(a): $L_{\rm 5~GHz}$ vs. $L_{\rm 2-10keV}$, where $L_{\rm 5~GHz}$ is the core radio luminosity at 5 GHz and $L_{\rm 2-10keV}$ is the X-ray luminosity in the 2--10 keV band. The black and red solid lines are the best linear-fitting lines for FR I and FR II RGs, respectively, with the corresponding 95\% confidence bands also shown. \emph{Panel}-(b): Same to \emph{Panel}-(a), however, this panel presents the core radio luminosity of CSOs at 8 GHz.}
\label{Lx_LR}
\end{figure*}

As illustrated in Figure \ref{Lx_LR}(a), $L_{\rm 2-10~keV}$ represents the 2--10 keV X-ray luminosity, while $L_{\rm 5~GHz}$ denotes the core radio luminosity at 5 GHz. We find that nearly all CSOs fall within the 95\% confidence band of the best linear fit for FR I RGs and are  distinctly separated from the FR II RGs. Given that a larger portion of CSOs in our sample have 8 GHz data available in the RFC, we use the core radio luminosity at 8 GHz instead of 5 GHz, as displayed in Figure \ref{Lx_LR}(b). The distribution of CSOs remains consistent with that observed in Figure \ref{Lx_LR}(a); the X-ray-to-radio luminosity ratios in CSOs closely align with those observed in FR I RGs. Notably, on average, CSOs exhibit higher luminosities than FR I RGs and are positioned at the high-luminosity extension of the 95\% confidence band of the best linear fit for FR I RGs. While the possibility of disk-corona contributions to X-ray emission in individual CSOs cannot be entirely ruled out, the X-ray emission of most CSOs in our sample is predominantly attributed to non-thermal processes associated with jets or lobes, similar to those observed in FR I RGs.

\subsubsection{$\Gamma_{\rm X}$ vs. $R_{\rm Edd}$}

The X-ray emission in RQ-AGNs is believed to originate from the IC process of the optical/UV photons by hot or relativistic electrons existing in the corona (\citealp{2000PASP..112.1145F, 2009A&ARv..17...47T}). A high Eddington ratio leads to an augmented IC cooling of the disc-corona system, and thus a steeper X-ray power-law spectrum (\citealp{2015A&ARv..23....1B} for a review). A positive correlation between $\Gamma_{\rm X}$ and $R_{\rm Edd}$ is expected and observed in RQ-AGNs (\citealp{2004ApJ...607L.107W, 2006ApJ...646L..29S, 2008ApJ...682...81S, 2009MNRAS.394..207C, 2009ApJ...700L...6R, 2019MNRAS.490.3793L}). Furthermore, the hard X-ray (2--10 keV) power-law spectral slope can serve as an estimator of the Eddington ratio in RQ-AGNs (e.g., \citealp{2008ApJ...682...81S, 2009ApJ...700L...6R}). Conversely, at low Eddington ratio ($R_{\rm Edd}<10^{-3}$), a negative correlation between $\Gamma_{\rm X}$ and $R_{\rm Edd}$ has been observed in low-luminosity AGNs (\citealp{2009MNRAS.399..349G, 2011A&A...530A.149Y, 2015MNRAS.447.1692Y, 2018ApJ...859..152S, 2023A&A...669A.114D}). These low-luminosity AGNs may have an accretion geometry distinct from that of luminous AGNs, characterized by advection-dominated accretion flows (ADAFs; \citealp{2014ARA&A..52..529Y}).

\begin{figure*}
 \centering
   \includegraphics[angle=0,scale=0.5]{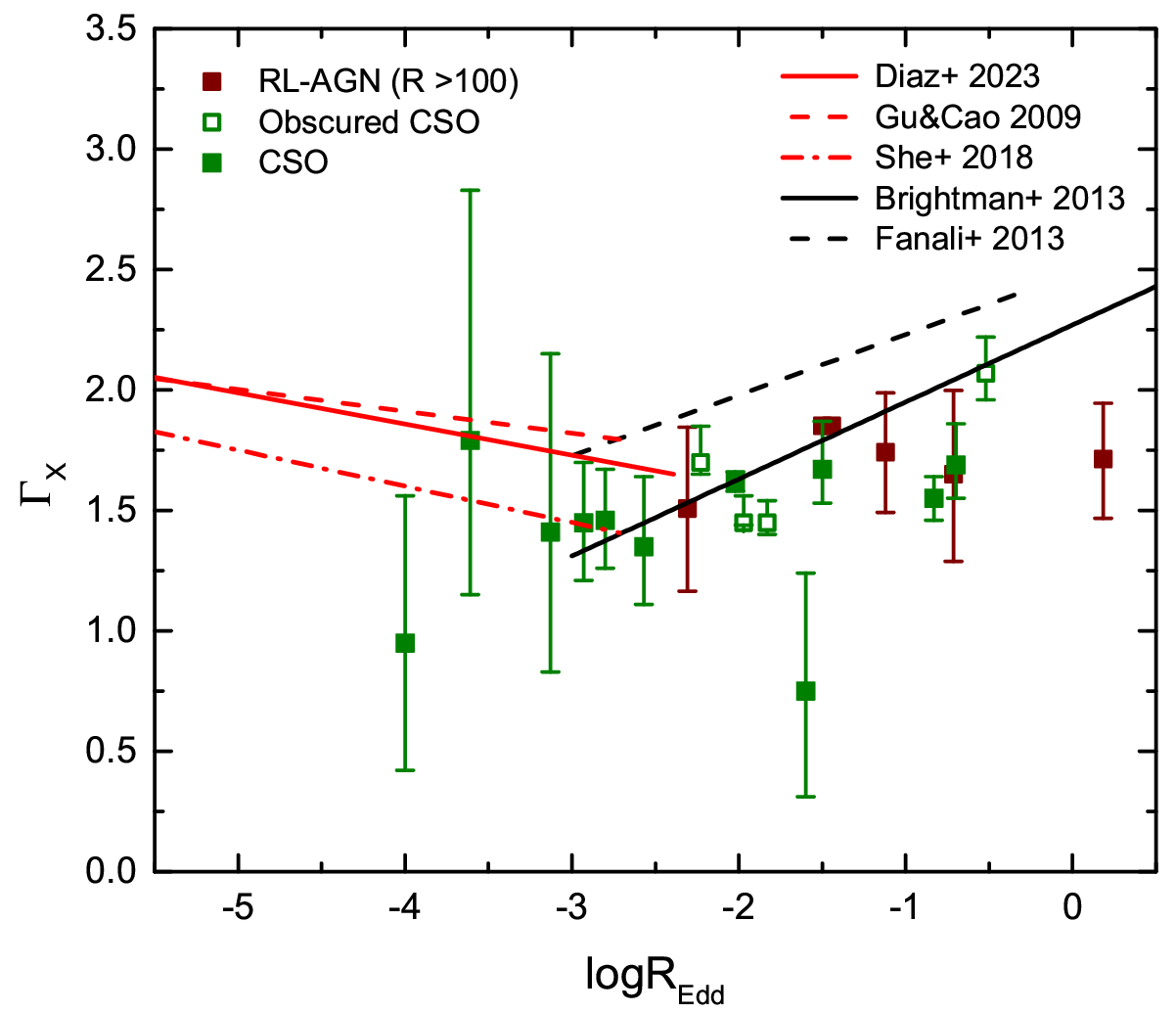}
\caption{$\Gamma_{\rm X}$ as a function of $R_{\rm Edd}$. The green solid squares represent the CSOs analyzed in this work, while the green opened squares represent the 4 obscured CSOs. The wine red solid squares indicate the very RL sources with $R > 100$ taken from \cite{2013MNRAS.433.2485B}. The black solid and dashed lines indicate the fitting lines for the RQ-AGN samples in \cite{2013MNRAS.433.2485B} and \cite{2013MNRAS.433..648F}, respectively. The red solid, dashed, and dash-dotted lines correspond to the fitting lines for the low-luminosity AGN samples from \cite{2023A&A...669A.114D}, \cite{2009MNRAS.399..349G}, and \cite{2018ApJ...859..152S}, respectively.}
\label{Gamm-Redd}
\end{figure*}

We collect the $R_{\rm Edd}$ values of CSOs in our sample from the literature, which are presented in Table \ref{tab-22CSO}. The corresponding references and calculation methods for $R_{\rm Edd}$ are also detailed in Table \ref{tab-22CSO}. It is important to acknowledge that we are currently unable to quantify the uncertainty in $R_{\rm Edd}$ arising from the different calculation methods. The relationship between $\Gamma_{\rm X}$ and $R_{\rm Edd}$ for CSOs is illustrated in Figure \ref{Gamm-Redd}. Considering the large errors of data points in Figure \ref{Gamm-Redd}, we employ the bootstrap method to sample within the error range (\citealp{MR0515681}) and estimate the correlation coefficient ($r$) between $\Gamma_{\rm X}$ and $R_{\rm Edd}$. This estimation yields a value of $r=0.27\pm0.26$. When considering only sources with $R_{\rm Edd} > 10^{-3}$, the correlation analysis results in $r=0.35\pm0.18$. These findings suggest that there is no significant correlation between $\Gamma_{\rm X}$ and $R_{\rm Edd}$ for these CSOs. This differs from the characteristics observed in RQ-AGNs. \cite{2020MNRAS.497..482L} investigated the relation between $\Gamma_{\rm X}$ and $R_{\rm Edd}$ in a young radio source sample but failed to identify any positive correlation. Similarly, \cite{2019MNRAS.490.3793L} reported no significant correlation existing between $\Gamma_{\rm X}$ and $R_{\rm Edd}$ for RL-AGNs while it was observed in an RQ-AGN sample. However, \cite{2013MNRAS.433.2485B} suggested that the RL-AGNs in their sample are generally consistent with the correlation trend observed in their RQ-AGN sample. Nevertheless, we observe that only considering the very RL sources with $R > 100$ ($R$: the radio-loudness parameter defined in \citealp{1989AJ.....98.1195K}), they do not obey this correlation in their figure 9. For comparative purposes, the RL sources with $R > 100$ in \cite{2013MNRAS.433.2485B}, along with the fitting line of the $\Gamma_{\rm X}$--$R_{\rm Edd}$ relationship for their RQ-AGN sample, are also present in Figure \ref{Gamm-Redd}. Additionally, we include the fitting lines between $\Gamma_{\rm X}$ and $R_{\rm Edd}$ reported in other studies (\citealp{2009MNRAS.399..349G, 2013MNRAS.433..648F, 2018ApJ...859..152S, 2023A&A...669A.114D}). It is observed that these CSOs, together with the RL-AGNs in \cite{2013MNRAS.433.2485B}, do not adhere to those established fitting lines. These findings further suggest that jet radiation, including both nucleus-jet and mini-lobe emission, should be regarded as the predominant source of X-ray emission in these CSOs.

\section{Summary} \label{sec:summ}

We have conducted a comprehensive analysis of the \chandra\, and \xmm\, X-ray observations for a sample of 22 CSOs, successfully deriving reliable X-ray spectra for 17 CSOs. Among these, 2 CSOs exhibit an iron emission line component, while 3 CSOs require an additional soft X-ray component to better fit their spectral characteristics. Multiple observations are available for 9 CSOs; however, none of them displays significant flux variation. Although most CSOs in our sample required an absorbing column density higher than the Galactic value to adequately model their X-ray spectra, the distribution of the derived $N_{\rm H}^{\rm int}$ values suggests that these CSOs are not heavily absorbed. 

We compiled a sample of $N_{\rm H}^{\rm int}$ values for FR I and FR II RGs from \cite{2004ApJ...617..915D}, \cite{2006ApJ...642...96E}, and \cite{2009MNRAS.396.1929H}. Comparing this with our CSO sample, we found that CSOs tend to have a similar $N_{\rm H}^{\rm int}$ distribution to FR I RGs. Additionally, the $\Gamma_{\rm X}$ distribution of CSOs is more closely aligned with that of RLQs and LERGs. In the radio--X-ray luminosity diagram, CSOs exhibit a distribution similar to that of FR I RGs, distinctly separated from FR II RGs, positioned at the high-luminosity extension of the 95\% confidence band of the best linear fit for FR I RGs. We therefore suggest that the jet radiation, encompassing both nucleus-jet and mini-lobe emissions, is the predominant source of X-ray emission in these CSOs. However, contributions from a disk-corona system cannot be entirely ruled out, especially for individual CSOs.

\acknowledgments

We thank the referee for the valuable suggestions that improved the manuscript. We thank the kind permission for the usage of the Radio Fundamental Catalog (RFC) data available at \url{https://astrogeo.smce.nasa.gov/rfc/}. This work is supported by the National Key R\&D Program of China (grant 2023YFE0117200) and the National Natural Science Foundation of China (grants 12203022, 12022305, 11973050).

\begin{deluxetable*}{lcccccc}
\tabletypesize{\scriptsize}
\tablecolumns{6} 
\tablewidth{0pc}
\tablecaption{Source information. Column (1) source name; Column (2) redshift; Column (3) angle size (in mas); Column (4) linear size (in kpc); Column (5) Eddington ratio; Column (6) references.}\tablenum{1}
\tablehead{\colhead{Source} & \colhead{$z^{\blacklozenge}$} & \colhead{Ang.Size$^{\blacklozenge}$} & \colhead{Lin.Size$^{\blacklozenge}$} & \colhead{log$R_{\rm Edd}$} & \colhead{Ref.$^{\clubsuit}$}\\
\colhead{(1)} & \colhead{(2)} & \colhead{(3)} & \colhead{(4)} & \colhead{(5)} & \colhead{(6)}} 
\startdata
22 CSOs\\
\hline
B2 0026+34 & 0.517 & 29.1 & 0.180 & -2.8 & [1] \\
0108+388 & 0.66847 & 8.0 & 0.056 & -1.31 & [2] \\
B2 0116+31 & 0.0602 & 100.0 & 0.115 & -1.94 & [2] \\
TXS 0128+554 & 0.03649 & 23.0 & 0.016 & - & - \\
B3 0402+379 & 0.05505 & 42.0 & 0.044 & - & - \\
B3 0710+439 & 0.518 & 35.0 & 0.217 & -0.83 & [2] \\
JVAS J1035+5628 & 0.46 & 38.0 & 0.221 & -3.38 & [1] \\
PKS 1117+146 & 0.362 & 101.0 & 0.507 & -3.13 & [1] \\
NGC 3894 & 0.01075 & 54.8 & 0.012 & -4 & [3] \\
JVAS J1234+4753 & 0.373082 & 27.4 & 0.140 & - & - \\
JVAS J1247+6723 & 0.107219 & 5.0 & 0.010 & -3.92 & [1] \\
DA 344 & 0.36801 & 68.0 & 0.345 & -1.5 & [2] \\
PKS B1345+125 & 0.121 & 100.0 & 0.215 & -2.93 & [1] \\
1358+625 & 0.431 & 67.6 & 0.378 & -2.43 & [1] \\
B3 1441+409 & 0.15 & 123.4 & 0.319 & - & - \\
4C 52.37 & 0.105689 & 250.0 & 0.478 & -3.61 & [1] \\
CTD 93 & 0.473 & 61.3 & 0.362 & -2.57 & [1] \\
PKS 1718--649 & 0.01443 & 7.0 & 0.002 & -2.02 & [2] \\
PKS 1934--63 & 0.183 & 42.6 & 0.130 & -0.7 & [2] \\
S4 1943+54 & 0.263 & 48.8 & 0.196 & - & - \\
S5 1946+70 & 0.101 & 40.6 & 0.075 & - & - \\
TXS 2352+495 & 0.23831 & 90.0 & 0.337 & -1.6 & [2] \\
\hline
4 obscured CSOs\\
\hline
OQ 208 & 0.077 & 11.0 & 0.016 & -1.97 & [1] \\
JVAS J1511+0518 & 0.084 & 10.6 & 0.017 & -2.23 & [1] \\
S4 2021+61 & 0.2266 & 29.0 & 0.104 & -1.83 & [2] \\
NGC 7674 & 0.02892 & 1300.0 & 0.744 & -0.52 & [4] \\
\enddata      
\tablenotetext{\blacklozenge}{The redshift, linear size (LS), and angular size of the 26 CSOs, excluding B3 1441+409, are taken from \cite{2024ApJ...961..240K}. The redshift and angular size of B3 1441+409 are obtained from Fanti et al. (2011) and \cite{2024ApJ...961..240K}, respectively, while its linear size is estimated by us.}
\tablenotetext{\clubsuit}{$R_{\rm Edd}=L_{\rm bol}/L_{\rm Edd}$, where $L_{\rm bol}$ is the bolometric luminosity and $L_{\rm Edd}=1.38 \times 10^{38}M_{\rm BH}/M_{\odot}$ is the Eddington luminosity. 
[1] \cite{2020MNRAS.491...92L}: $L_{\rm bol}$ is estimated from the luminosity of emission lines H$\beta$ or [OIII]. $M_{\rm BH}$ is estimated using the line width and luminosity of the broad line H$\beta$ through an empirical relation, or by the relation between $M_{\rm BH}$ and stellar velocity dispersions $\sigma_{\ast}$;
[2] \cite{2020ApJ...892..116W}: $M_{\rm BH}$ is estimated using the bulge luminosity-$M_{\rm BH}$ relation, or host galaxy absolute magnitude in R-band with the $M_{\rm R}$-$M_{\rm BH}$ relation, or the $\sigma_{\ast}$-$M_{\rm BH}$ relation; $L_{\rm bol}$ is also estimated from the luminosity of emission lines H$\beta$ or [OIII];
[3] \cite{2021ApJ...922...84B}: $M_{\rm BH}$ is estimated using the $\sigma_{\ast}$-$M_{\rm BH}$ relation. The upper limit on accretion disk luminosity, estimated from the observed IR flux, is considered as $L_{\rm bol}$;
[4] \cite{2002ApJ...579..530W}: $M_{\rm BH}$ is estimated using the $\sigma_{\ast}$-$M_{\rm BH}$ relation. A bolometric correction with a range of $L_{\rm bol}/L_{2-10 \rm keV} \approx 10-20$ is used to estimate $L_{\rm bol}$.}
\label{tab-22CSO}
\end{deluxetable*}

\begin{deluxetable*}{lcccccc}
\tabletypesize{\scriptsize} 
\tablecolumns{6} 
\tablewidth{0pc}
\tablecaption{\chandra\, observations. Column (1) source name; Column (2) observation date; Column (3) observation ID; Column (4) observation exposure time (in ks); Column (5) total counts within the range of 0.5 to 7.5 keV in the source extracted region; Column (6) net counts within the range of 0.5 to 7.5 keV in the source extracted region.}\tablenum{2}
\tablehead{\colhead{Source} & \colhead{Date} & \colhead{Obs.ID} & \colhead{Exposure} & \colhead{Total} & \colhead{Net} \\
\colhead{(1)} & \colhead{(2)} & \colhead{(3)} & \colhead{(4)} & \colhead{(5)} & \colhead{(6)}}
\startdata
B2 0116+31&2010-11-04&12848&4.74&7&5\\
TXS	0128+554&2019-03-29&21408&5.73&438&437\\
            &2019-03-30&22160&6.68&565&483\\
            &2019-03-31&22161&3.84&327&279\\
            &2019-04-01&22162&3.06&278&238\\
B3 0402+379&2011-04-04&12704&9.99&414&222\\
           &2013-11-06&16120&94.74&4147&2355\\
B3 0710+439&2011-01-18&12845&37.85&1781&1763\\
NGC	3894&2009-07-20&10389&38.54&510&495\\
JVAS J1234+4753$^{\blacklozenge}$&2002-03-23&3055&4.76&-&-\\
PKS	B1345+125&2000-02-24&836&25.35&1461&1427\\
4C 52.37&2007-06-03&8257&19.91&46&39\\
CTD	93&2010-12-04&12846&37.85&225&202\\
PKS 1718--649$^{\blacklozenge}$&2010-11-09&12849&4.78&238&232\\
            &2014-06-20&16070&15.94&1233&1219\\
            &2014-06-23&16623&33.04&-&-\\
PKS	1934--63&2010-07-08&11504&19.79&392&385\\
S4 1943+54&2011-05-04&12851&4.78&13&11\\
S5 1946+70&2011-02-06&12852&4.74&119&118\\
\enddata
\tablenotetext{\blacklozenge}{The observations of JVAS J1234+4753 and PKS 1718--649 (Obs-ID 16623) are affected by the pileup effect.}
\label{tab-chandra}
\end{deluxetable*}

\begin{deluxetable*}{lcccccc}
\tabletypesize{\scriptsize} 
\tablecolumns{6} 
\tablewidth{0pc}
\tablecaption{\xmm\, observations. Column (1) source name; Column (2) observation date; Column (3) observation ID; Column (4) clean exposure time (in ks) after excluding the period of flaring high background activity for PN and MOS; Column (5) total counts within the range of 0.5 to 7.5 keV in the source extracted region for PN and MOS; Column (6) net counts within the range of 0.5 to 7.5 keV in the source extracted region for PN and MOS.}\tablenum{3}
\tablehead{\colhead{Source} & \colhead{Date} & \colhead{Obs.ID} & \colhead{Exposure  (PN/MOS)} & \colhead{Total (PN/MOS)} & \colhead{Net (PN/MOS)} \\
\colhead{(1)} & \colhead{(2)} & \colhead{(3)} & \colhead{(4)} & \colhead{(5)} & \colhead{(6)}}
\startdata
B2 0026+34&2004-01-08&0205180101&10.33/13.37&548/458&485/395\\
0108+388&2004-01-09&0202520101&13.37/16.40&126/72&53/13\\
B3 0710+439&2004-03-22&0202520201&11.79/14.52&1147/1043&1079/990\\
JVAS J1035+5628$^{\blacklozenge}$&2004-10-21&0202520301&1.24/6.95&34/114&22/60\\
PKS 1117+146&2007-06-13&0502510201&15.65/21.32&111/72&76/44\\
NGC 3894$^{\clubsuit}$&2022-11-14&0904530201&7.07/9.57&24/35&-/1\\
        &2022-11-14&0904530101&33.44/61.33&1784/1952&1377/1480\\
JVAS J1234+4753$^{\blacklozenge}$&2004-06-17&0205180301&-/1.37&-/263&-/254\\
JVAS J1247+6723&2005-11-29&0306680201&41.45/50.48&291/233&19/51\\
DA 344&2007-12-05&0502510301&19.83/24.56&774/577&642/468\\
1358+625$^{\blacklozenge}$&2004-04-14&0202520401&-/3.96&-/61&-/-\\
B3 1441+409&2019-01-18&0822530101&33.01/67.06&623/460&426/267\\
CTD 93$^{\blacklozenge}$&2008-01-17&0502510401&-/7.14&-/222&-/84\\
      &2008-01-19&0502510801&0.09/2.10&2/31&1/11\\
PKS 1718--649&2017-03-05&0784530201&13.69/27.46&2108/2266&1899/2073\\
            &2018-03-08&0804520301&29.29/38.14&4264/3456&3765/3162\\
            &2020-03-27&0845110101&108.60/3.75&16047/22&14603/-\\
PKS	1934--63&2017-04-01&0784610201&18.83/29.53&718/658&669/608\\
S5 1946+70&2016-10-21&0784610101&17.32/15.71&912/538&763/458\\
TXS 2352+495&2003-12-25&0202520501&7.58/17.43&134/234&75/110\\
\enddata
\tablenotetext{\blacklozenge}{Both the PN and MOS observations of JVAS J1035+5628, JVAS J1234+4753, and 1358+625 reveal that high background activities persisted throughout the entire observation period. Additionally, the PN observation of CTD 93 is consistently affected by high background bursts for the majority of the time, whereas the MOS observation of CTD 93 experiences partial effects from these bursts.}
\tablenotetext{\clubsuit}{The first observation time is from 2022-11-14 06:46:16 to 2022-11-14 09:56:36, during which the source is not detected. The second observation time is from 2022-11-14 09:56:36 to 2022-11-15 06:54:56.}
\label{tab-XMM}
\end{deluxetable*}

\begin{deluxetable*}{lcccccccc}
\tabletypesize{\scriptsize} 
\tablecolumns{8} 
\tablewidth{0pc}
\tablecaption{\chandra\, observations with pile-up effect. Column (1) source name; Column (2) observation date; Column (3) observation ID; Column (4) observation exposure time (in ks); Column (5) inner radius of the source extracted region (in arcsec); Column (6) outer radius of the source extracted region (in arcsec); Column (7) total counts within the range of 0.5 to 7.5 keV in the source extracted region; Column (8) net counts within the range of 0.5 to 7.5 keV in the source extracted region.}\tablenum{4}
\tablehead{\colhead{Source} & \colhead{Date} & \colhead{Obs.ID} & \colhead{Exposure} & \colhead{$r_{\rm i}$} & \colhead{$r_{\rm o}$} & \colhead{Total} & \colhead{Net}\\
\colhead{(1)} & \colhead{(2)} & \colhead{(3)} & \colhead{(4)} & \colhead{(5)} & \colhead{(6)} & \colhead{(7)} & \colhead{(8)}}
\startdata
JVAS J1234+4753&2002-03-23&3055&4.76&$\sim 1$&7&78&76\\
PKS 1718--649&2014-06-23&16623&33.04&$\sim 0.7$&7&822&789\\
\enddata
\label{tab-pileup}
\end{deluxetable*}

\begin{deluxetable*}{lccccccccccc}
\tabletypesize{\scriptsize} 
\tablecolumns{10} 
\tablewidth{0pc}
\tablecaption{Joint spectral fitting results. Column (1) source name; Column (2)
observation; Column (3) fitting model; Column (4) intrinsic column density (in 10$^{22}$ cm$^{-2}$); Column (5) photon spectral index; Column (6) norm of power-law model (in 10$^{-4}$ photons keV$^{-1}$ cm$^{-2}$ s$^{-1}$ at 1 keV); Column (7) plasma temperature (in keV); Column (8) energy of Gaussian line (in keV); Column (9) width of the Gaussian line (in keV); Column (10) flux of the power-law component in the 2--10 keV band (in 10$^{-13}$ erg cm$^{-2}$ s$^{-1}$), corrected for absorption; Column (11) fitting statistics.}\tablenum{5}
\tablehead{\colhead{Source$^{\blacklozenge}$} & \colhead{Obs} & \colhead{Model$^{\clubsuit}$} & \colhead{$N_{\rm H}^{\rm int}$} & \colhead{$\Gamma_{\rm X}$} & \colhead{$A$} & \colhead{$kT$} & \colhead{$E_\mathrm{l}$} & \colhead{$\sigma_\mathrm{l}$} & \colhead{$F_{\rm 2-10 keV}$} & \colhead{$\chi^2$/d.o.f}\\
\colhead{(1)} & \colhead{(2)} & \colhead{(3)} & \colhead{(4)} & \colhead{(5)} & \colhead{(6)} & \colhead{(7)} & \colhead{(8)} & \colhead{(9)} & \colhead{(10)} & \colhead{(11)}}
\startdata
TXS	0128+554&\chandra\,&PL&0.34$_{-0.19}^{+0.21}$&2.35$_{-0.24}^{+0.25}$&4.26$_{-1.09}^{+1.51}$&-&-&-&6.55$_{-0.82}^{+0.79}$&47.18/65\\
B3 0402+379&\chandra\,&PL+zgauss&0.71$_{-0.15}^{+0.17}$&2.96$_{-0.21}^{+0.23}$&2.30$_{-0.58}^{+0.78}$&-&6.57$_{-0.05}^{+0.05}$&0.01 (fix)&1.55$_{-0.31}^{+0.28}$&157.91/153\\
B3 0710+439&\chandra\,/\xmm\,&PL&0.43$_{-0.10}^{+0.11}$&1.55$_{-0.09}^{+0.09}$&0.87$_{-0.08}^{+0.09}$&-&-&-&4.52$_{-0.33}^{+0.32}$&148.94/171\\
S5 1946+70&\chandra\,/\xmm\,&PL&1.16$_{-0.39}^{+0.51}$&0.95$_{-0.23}^{+0.25}$&0.43$_{-0.14}^{+0.22}$&-&-&-&6.00$_{-1.18}^{+1.03}$&90.81/70\\
PKS 1718--649&\chandra\,&PL&0.20$_{-0.06}^{+0.06}$&1.92$_{-0.12}^{+0.13}$&1.12$_{-0.17}^{+0.20}$&-&-&-&3.31$_{-0.46}^{+0.46}$&130.14/95\\
PKS 1718--649&\xmm\,&APEC+PL&0.03$_{-0.01}^{+0.01}$&1.62$_{-0.04}^{+0.04}$&0.81$_{-0.05}^{+0.05}$&0.80$_{-0.04}^{+0.05}$&-&-&3.75$_{-0.18}^{+0.18}$&921.76/994\\
\enddata
\tablenotetext{\blacklozenge}{For PKS 1718--649, the spectra observed with \xmm\ require an additional soft X-ray component for a more accurate fit, whereas a simple absorbed power-law model adequately explains the \chandra\ observations. Therefore, we separately combine the multiple \chandra\, and \xmm\, observations to constrain the $N_{\rm H}^{\rm int}$ value.}
\tablenotetext{\clubsuit}{PL---power-law model; APEC---plasma model; zgauss---Gaussian model for emission line. Metal abundances in APEC model are fixed at solar value.}
\label{Joint-fit}
\end{deluxetable*}

\begin{deluxetable*}{lccccccccccc}
\tabletypesize{\scriptsize} 
\tablecolumns{11} 
\tablewidth{0pc}
\tablecaption{The \chandra\, and \xmm\, observation analysis results. Column (1) source name; Column (2) observation date; Column (3) observation ID; Column (4) fitting model; Column (5) intrinsic column density (in 10$^{22}$ cm$^{-2}$); Column (6) photon spectral index; Column (7) norm of power-law model (in 10$^{-5}$ photons keV$^{-1}$ cm$^{-2}$ s$^{-1}$ at 1 keV); Column (8) plasma temperature (in keV); Column (9) energy of Gaussian line (in keV); Column (10) flux of the power-law component in the 2--10 keV band (in 10$^{-13}$ erg cm$^{-2}$ s$^{-1}$), corrected for absorption; Column (11) fitting statistics.}\tablenum{6}
\tablehead{\colhead{Source$^{\blacklozenge}$} & \colhead{Date} & \colhead{Obs.ID} & \colhead{Model$^{\clubsuit}$} & \colhead{$N_{\rm H}^{\rm int}$} & \colhead{$\Gamma_{\rm X}$$^{\spadesuit}$} & \colhead{$A$} & \colhead{$kT$} & $E_\mathrm{l}$ & \colhead{$F_{\rm 2-10 keV}$} & \colhead{Fit Statistic}\\
\colhead{(1)} & \colhead{(2)} & \colhead{(3)} & \colhead{(4)} & \colhead{(5)} & \colhead{(6)} & \colhead{(7)} & \colhead{(8)} & \colhead{(9)} & \colhead{(10)} & \colhead{(11)}}
\startdata
\chandra\\
\hline
TXS	0128+554&2019-03-29&21408&PL&0.34 (fix)&2.42$_{-0.21}^{+0.22}$&44.50$_{-7.02}^{+7.63}$&-&-&6.22$_{-1.15}^{+1.06}$&13.92/17\\
            &2019-03-30&22160&PL&0.34 (fix)&2.27$_{-0.19}^{+0.20}$&38.89$_{-6.24}^{+6.76}$&-&-&6.77$_{-1.15}^{+1.05}$&13.27/23\\
            &2019-03-31&22161&PL&0.34 (fix)&2.31$_{-0.23}^{+0.25}$&38.56$_{-7.76}^{+8.70}$&-&-&6.35$_{-1.36}^{+1.17}$&14.97/13\\
            &2019-04-01&22162&PL&0.34 (fix)&2.49$_{-0.33}^{+0.35}$&49.48$_{-12.13}^{+14.24}$&-&-&6.31$_{-1.82}^{+1.59}$&3.69/10\\
B3 0402+379&2011-04-04&12704&PL&0.71 (fix)&2.60$_{-0.46}^{+0.52}$&18.55$_{-6.05}^{+7.46}$&-&-&2.01$_{-0.86}^{+0.70}$&9.01/15\\
           &2013-11-06&16120&PL+zgauss&0.71 (fix)&2.97$_{-0.11}^{+0.12}$&27.94$_{-2.30}^{+2.40}$&-&6.57 (fix) 
           &1.86$_{-0.18}^{+0.17}$&147.56/139\\
B3 0710+439&2011-01-18&12845&PL&0.43 (fix)&1.49$_{-0.07}^{+0.07}$&8.41$_{-0.52}^{+0.52}$&-&- 
            &4.76$_{-0.40}^{+0.39}$&76.14/71\\
NGC 3894&2009-07-20&10389&APEC+PL+zgauss&1.58$_{-0.88}^{+1.18}$&0.95$_{-0.53}^{+0.61}$&1.75$_{-0.97}^{+2.71}$&0.85$_{-0.09}^{+0.11}$&6.42$_{-0.10}^{+0.11}$&2.47$_{-0.44}^{+0.42}$&88.6/79\\
JVAS J1234+4753&2002-03-23&3055&PL&<0.44&2.45$_{-0.47}^{+0.96}$&34.47$_{-7.93}^{+32.51}$&-&- 
            &4.68$_{-3.45}^{+1.94}$&9.36/11\\
PKS B1345+125&2000-02-24&836&APEC+PL&3.20$_{-0.56}^{+0.63}$&1.45$_{-0.24}^{+0.25}$&22.77$_{-7.03}^{+10.88}$&0.83$_{-0.10}^{+0.10}$&-&13.80$_{-0.96}^{+1.00}$&79.76/61\\
4C 52.37&2007-06-03&8257&PL&<0.31&1.79$_{-0.64}^{+1.04}$&0.29$_{-0.10}^{+0.30}$&-&-&0.10$_{-0.03}^{+0.03}$&14.8/5\\
CTD	93&2010-12-04&12846&PL&<0.32&1.35$_{-0.24}^{+0.29}$&0.65$_{-0.12}^{+0.21}$&-&- 
            &0.47$_{-0.07}^{+0.06}$&11.11/7\\
PKS 1718--649&2010-11-09&12849&PL&0.20 (fix)&1.85$_{-0.26}^{+0.28}$&10.82$_{-1.98}^{+2.07}$&-&-
            &3.47$_{-1.04}^{+0.96}$&4.59/9\\
            &2014-06-20&16070&PL&0.20 (fix)&1.70$_{-0.09}^{+0.09}$&17.20$_{-1.30}^{+1.33}$&-&- 
            &7.08$_{-0.65}^{+0.65}$&44.73/50\\
            &2014-06-23&16623&PL&0.20 (fix)&2.25$_{-0.12}^{+0.12}$&37.86$_{-3.37}^{+3.42}$&-&- 
            &6.76$_{-0.93}^{+0.78}$&45.10/35\\
PKS	1934--63&2010-07-08&11504&PL&<0.10&1.65$_{-0.16}^{+0.19}$&2.99$_{-0.36}^{+0.58}$&-&- 
            &1.33$_{-0.28}^{+0.25}$&18.84/15\\
S4 1943+54&2011-05-04&12851&PL&<3.26&1.7 (fix)&0.60$_{-0.30}^{-0.86}$&-&- 
            &0.25$_{-0.16}^{+0.12}$&1.55/4\\
S5 1946+70&2011-02-06&12852&PL&1.16 (fix)&1.50$_{-0.36}^{+0.36}$&8.06$_{-2.52}^{+3.44}$&-&-&4.51$_{-1.32}^{+1.31}$&11.85/19\\
\hline
\xmm\\
\hline
B2 0026+34&2004-01-08&0205180101&PL&0.85$_{-0.31}^{+0.36}$&1.46$_{-0.20}^{+0.21}$&4.11$_{-0.92}^{+1.22}$&-&-&2.44$_{-0.35}^{+0.32}$&48.43/44\\
B3 0710+439&2004-03-22&0202520201&PL&0.43 (fix)&1.59$_{-0.07}^{+0.07}$&7.56$_{-0.52}^{+0.53}$&-&- 
          &3.67$_{-0.33}^{+0.31}$&70.33/100\\
PKS 1117+146&2007-06-13&0502510201&PL&<0.24&1.41$_{-0.58}^{+0.74}$&0.25$_{-0.09}^{+0.13}$&-&-&0.16$_{-0.06}^{+0.04}$&10.9/6\\
NGC 3894&2022-11-14&0904530101&APEC+PL&<0.02&0.85$_{-0.10}^{+0.10}$&1.07$_{-0.13}^{+0.14}$&0.36$_{-0.04}^{+0.07}$&-
&1.77$_{-0.12}^{+0.13}$&225.71/167\\
JVAS J1234+4753&2004-06-17&0205180301&PL&<0.19&2.21$_{-0.23}^{+0.44}$&21.28$_{-2.50}^{+7.73}$&-&- 
            &4.02$_{-1.35}^{+1.19}$&12/9\\
DA 344&2007-12-05&0502510301&PL&<0.12&1.67$_{-0.14}^{+0.20}$&1.88$_{-0.21}^{+0.37}$&-&-&0.81$_{-0.13}^{+0.12}$&42.62/59\\
B3 1441+409&2019-01-18&0822530101&PL&0.16$_{-0.11}^{+0.13}$&1.85$_{-0.31}^{+0.36}$&2.29$_{-0.59}^{+0.82}$&-&-&0.74$_{-0.23}^{+0.20}$&59.62/48\\
CTD 93&2008-01-17&0502510401&PL&<2.73&2.24$_{-0.75}^{+3.87}$&1.70$_{-0.61}^{+21.95}$&-&-&0.31$_{-0.11}^{+0.08}$&3.71/7\\
PKS 1718--649&2017-03-05&0784530201&APEC+PL&0.03 (fix)&1.54$_{-0.07}^{+0.07}$&7.73$_{-0.60}^{+0.62}$&0.8 (fix)&-
          &4.07$_{-0.30}^{+0.29}$&179.64/198\\
          &2018-03-08&0804520301&APEC+PL&0.03 (fix)&1.61$_{-0.05}^{+0.05}$&7.80$_{-0.44}^{+0.44}$&0.8 (fix)&-&3.66$_{-0.20}^{+0.19}$&286.05/322\\
          &2020-03-27&0845110101&APEC+PL&0.03 (fix)&1.65$_{-0.04}^{+0.04}$&8.31$_{-0.28}^{+0.28}$&0.8 (fix)&-&3.66$_{-0.13}^{+0.12}$&448.52/472\\
PKS	1934--63&2017-04-01&0784610201&PL&<0.10&1.69$_{-0.14}^{+0.17}$&2.75$_{-0.36}^{+0.47}$&-&- 
          &1.14$_{-0.16}^{+0.15}$&86.38/61\\
S5 1946+70&2016-10-21&0784610101&PL&1.16 (fix)&0.91$_{-0.11}^{+0.11}$&3.13$_{-0.45}^{+0.49}$&-&- 
          &4.67$_{-0.45}^{+0.43}$&84.46/67\\
TXS 2352+495&2003-12-25&0202520501&PL&<0.36&0.75$_{-0.44}^{+0.49}$&0.43$_{-0.16}^{+0.24}$&-&- 
          &0.85$_{-0.31}^{+0.22}$&19.72/13\\
\enddata
\tablenotetext{\blacklozenge}{For NGC 3894, when conducting joint fitting of the \chandra\, and \xmm\, spectra with a minimum of 20 counts per energy bin, the $N_{\rm H}^{\rm int}$ value remains unconstrained, resulting in a reduced $\chi^2$ value of 1.6. Furthermore, fixing the $N_{\rm H}^{\rm int}$ value based on the \chandra\, spectrum fit and applying it to the \xmm\, spectrum yields a reduced $\chi^2$ value of 1.8. Therefore, we opted to fit the \chandra\, and \xmm\, spectra independently. Additionally, a minimum of 20 counts per energy bin may smooth out emission line signals; therefore, for the \chandra\ spectrum, a minimum of 5 counts per energy bin is finally set, and the C-statistic method is employed.\\
For JVAS J1234+4753, CTD 93, and PKS 1934--63, the joint spectral fitting remains inconclusive for determining the value of $N_{\rm H}^{\rm int}$, we thus separately fit their \chandra\, and \xmm\, spectra.}
\tablenotetext{\clubsuit}{Metal abundances in APEC model are fixed at solar value. The plasma temperature in the APEC model, as well as the energy and width of the Gaussian line, are also fixed to the values obtained from joint spectral fitting.}
\tablenotetext{\spadesuit}{$\Gamma_{\rm X}$ is fixed at 1.7 in the X-ray spectrum fitting of S4 1943+54 due to insufficient statistical data.}
\label{tab-Result}
\end{deluxetable*}

\clearpage
\bibliography{reference}{}

\begin{thebibliography}{}
\expandafter\ifx\csname natexlab\endcsname\relax\def\natexlab#1{#1}\fi
\providecommand{\url}[1]{\href{#1}{#1}}
\providecommand{\dodoi}[1]{doi:~\href{http://doi.org/#1}{\nolinkurl{#1}}}
\providecommand{\doeprint}[1]{\href{http://ascl.net/#1}{\nolinkurl{http://ascl.net/#1}}}
\providecommand{\doarXiv}[1]{\href{https://arxiv.org/abs/#1}{\nolinkurl{https://arxiv.org/abs/#1}}}

\bibitem[{{An} {et~al.}(2012){An}, {Wu}, {Yang}, {Taylor}, {Hong}, {Baan}, {Liu}, {Wang}, {Zhang}, {Wang}, {Chen}, {Cui}, {Hao}, \& {Zhu}}]{2012ApJS..198....5A}
{An}, T., {Wu}, F., {Yang}, J., {et~al.} 2012, \apjs, 198, 5, \dodoi{10.1088/0067-0049/198/1/5}

\bibitem[{{Balasubramaniam} {et~al.}(2021){Balasubramaniam}, {Stawarz}, {Cheung}, {Sobolewska}, {Marchenko}, {Thimmappa}, {Kr{\'o}l}, {Migliori}, \& {Siemiginowska}}]{2021ApJ...922...84B}
{Balasubramaniam}, K., {Stawarz}, {\L}., {Cheung}, C.~C., {et~al.} 2021, \apj, 922, 84, \dodoi{10.3847/1538-4357/ac1ff5}

\bibitem[{{Balmaverde} {et~al.}(2006){Balmaverde}, {Capetti}, \& {Grandi}}]{2006A&A...451...35B}
{Balmaverde}, B., {Capetti}, A., \& {Grandi}, P. 2006, \aap, 451, 35, \dodoi{10.1051/0004-6361:20053799}

\bibitem[{{Belsole} {et~al.}(2006){Belsole}, {Worrall}, \& {Hardcastle}}]{2006MNRAS.366..339B}
{Belsole}, E., {Worrall}, D.~M., \& {Hardcastle}, M.~J. 2006, \mnras, 366, 339, \dodoi{10.1111/j.1365-2966.2005.09882.x}

\bibitem[{{Beuchert} {et~al.}(2018){Beuchert}, {Rodr{\'\i}guez-Ardila}, {Moss}, {Schulz}, {Kadler}, {Wilms}, {Angioni}, {Callingham}, {Gr{\"a}fe}, {Krau{\ss}}, {Kreikenbohm}, {Langejahn}, {Leiter}, {Maccagni}, {M{\"u}ller}, {Ojha}, {Ros}, \& {Tingay}}]{2018A&A...612L...4B}
{Beuchert}, T., {Rodr{\'\i}guez-Ardila}, A., {Moss}, V.~A., {et~al.} 2018, \aap, 612, L4, \dodoi{10.1051/0004-6361/201833064}

\bibitem[{{Bicknell} {et~al.}(1997){Bicknell}, {Dopita}, \& {O'Dea}}]{1997ASPC..121..672B}
{Bicknell}, C.~V., {Dopita}, M.~A., \& {O'Dea}, C.~P. 1997, in Astronomical Society of the Pacific Conference Series, Vol. 121, IAU Colloq. 163: Accretion Phenomena and Related Outflows, ed. D.~T. {Wickramasinghe}, G.~V. {Bicknell}, \& L.~{Ferrario}, 672

\bibitem[{{Brandt} \& {Alexander}(2015)}]{2015A&ARv..23....1B}
{Brandt}, W.~N., \& {Alexander}, D.~M. 2015, \aapr, 23, 1, \dodoi{10.1007/s00159-014-0081-z}

\bibitem[{{Brightman} {et~al.}(2013){Brightman}, {Silverman}, {Mainieri}, {Ueda}, {Schramm}, {Matsuoka}, {Nagao}, {Steinhardt}, {Kartaltepe}, {Sanders}, {Treister}, {Shemmer}, {Brandt}, {Brusa}, {Comastri}, {Ho}, {Lanzuisi}, {Lusso}, {Nandra}, {Salvato}, {Zamorani}, {Akiyama}, {Alexander}, {Bongiorno}, {Capak}, {Civano}, {Del Moro}, {Doi}, {Elvis}, {Hasinger}, {Laird}, {Masters}, {Mignoli}, {Ohta}, {Schawinski}, \& {Taniguchi}}]{2013MNRAS.433.2485B}
{Brightman}, M., {Silverman}, J.~D., {Mainieri}, V., {et~al.} 2013, \mnras, 433, 2485, \dodoi{10.1093/mnras/stt920}

\bibitem[{{Bronzini} {et~al.}(2024){Bronzini}, {Migliori}, {Vignali}, {Sobolewska}, {Stawarz}, {Siemiginowska}, {Orienti}, {D'Ammando}, {Giroletti}, {Principe}, \& {Balasubramaniam}}]{2024A&A...684A..65B}
{Bronzini}, E., {Migliori}, G., {Vignali}, C., {et~al.} 2024, \aap, 684, A65, \dodoi{10.1051/0004-6361/202348208}

\bibitem[{{Callingham} {et~al.}(2017){Callingham}, {Ekers}, {Gaensler}, {Line}, {Hurley-Walker}, {Sadler}, {Tingay}, {Hancock}, {Bell}, {Dwarakanath}, {For}, {Franzen}, {Hindson}, {Johnston-Hollitt}, {Kapi{\'n}ska}, {Lenc}, {McKinley}, {Morgan}, {Offringa}, {Procopio}, {Staveley-Smith}, {Wayth}, {Wu}, \& {Zheng}}]{2017ApJ...836..174C}
{Callingham}, J.~R., {Ekers}, R.~D., {Gaensler}, B.~M., {et~al.} 2017, \apj, 836, 174, \dodoi{10.3847/1538-4357/836/2/174}

\bibitem[{{Cao}(2009)}]{2009MNRAS.394..207C}
{Cao}, X. 2009, \mnras, 394, 207, \dodoi{10.1111/j.1365-2966.2008.14347.x}

\bibitem[{{Cao} {et~al.}(2024{\natexlab{a}}){Cao}, {Aharonian}, {An}, {Axikegu}, {Bai}, {Bao}, {Bastieri}, {Bi}, {Bi}, {Cai}, {Cao}, {Cao}, {Cao}, {Chang}, {Chang}, {Chen}, {Chen}, {Chen}, {Chen}, {Chen}, {Chen}, {Chen}, {Chen}, {Chen}, {Chen}, {Chen}, {Chen}, {Cheng}, {Cheng}, {Cui}, {Cui}, {Cui}, {Cui}, {Dai}, {Dai}, {Dai}, {Danzengluobu}, {Della Volpe}, {Dong}, {Duan}, {Fan}, {Fan}, {Fang}, {Fang}, {Feng}, {Feng}, {Feng}, {Feng}, {Feng}, {Gabici}, {Gao}, {Gao}, {Gao}, {Gao}, {Gao}, {Gao}, {Ge}, {Geng}, {Giacinti}, {Gong}, {Gou}, {Gu}, {Guo}, {Guo}, {Guo}, {Guo}, {Han}, {He}, {He}, {He}, {He}, {He}, {Heller}, {Hor}, {Hou}, {Hou}, {Hou}, {Hu}, {Hu}, {Hu}, {Huang}, {Huang}, {Huang}, {Huang}, {Huang}, {Huang}, {Huang}, {Ji}, {Jia}, {Jia}, {Jiang}, {Jiang}, {Jiang}, {Jin}, {Kang}, {Ke}, {Kuleshov}, {Kurinov}, {Li}, {Li}, {Li}, {Li}, {Li}, {Li}, {Li}, {Li}, {Li}, {Li}, {Li}, {Li}, {Li}, {Li}, {Li}, {Li}, {Li}, {Li}, {Li}, {Liang}, {Liang}, {Lin}, {Liu}, {Liu}, {Liu}, {Liu}, {Liu}, {Liu}, {Liu}, {Liu}, {Liu},
  {Liu}, {Liu}, {Liu}, {Liu}, {Liu}, {Lu}, {Luo}, {Lv}, {Ma}, {Ma}, {Ma}, {Mao}, {Min}, {Mitthumsiri}, {Mu}, {Nan}, {Neronov}, {Ou}, {Pang}, {Pattarakijwanich}, {Pei}, {Qi}, {Qi}, {Qiao}, {Qin}, {Ruffolo}, {S{\'a}iz}, {Semikoz}, {Shao}, {Shao}, {Shchegolev}, {Sheng}, {Shu}, {Song}, {Stenkin}, {Stepanov}, {Su}, {Sun}, {Sun}, {Sun}, {Tam}, {Tang}, {Tang}, {Tian}, {Wang}, {Wang}, {Wang}, {Wang}, {Wang}, {Wang}, {Wang}, {Wang}, {Wang}, {Wang}, {Wang}, {Wang}, {Wang}, {Wang}, {Wang}, {Wang}, {Wang}, {Wang}, {Wang}, {Wang}, {Wang}, {Wei}, {Wei}, {Wei}, {Wen}, {Wu}, {Wu}, {Wu}, {Wu}, {Wu}, {Xi}, {Xia}, {Xia}, {Xiang}, {Xiao}, {Xiao}, {Xin}, {Xin}, {Xing}, {Xiong}, {Xu}, {Xu}, {Xu}, {Xu}, {Xue}, {Yan}, {Yan}, {Yan}, {Yang}, {Yang}, {Yang}, {Yang}, {Yang}, {Yang}, {Yang}, {Yang}, {Yang}, {Yao}, {Yao}, {Ye}, {Yin}, {Yin}, {You}, {You}, {Yu}, {Yuan}, {Yue}, {Zeng}, {Zeng}, {Zeng}, {Zha}, {Zhang}, {Zhang}, {Zhang}, {Zhang}, {Zhang}, {Zhang}, {Zhang}, {Zhang}, {Zhang}, {Zhang}, {Zhang}, {Zhang}, {Zhang}, {Zhang}, {Zhang},
  {Zhang}, {Zhang}, {Zhang}, {Zhao}, {Zhao}, {Zhao}, {Zhao}, {Zhao}, {Zheng}, {Zhou}, {Zhou}, {Zhou}, {Zhou}, {Zhou}, {Zhou}, {Zhou}, {Zhu}, {Zhu}, {Zhu}, {Zhu}, {Zuo}, \& {(The Lhaaso Collaboration)}}]{2024ApJS..271...25C}
{Cao}, Z., {Aharonian}, F., {An}, Q., {et~al.} 2024{\natexlab{a}}, \apjs, 271, 25, \dodoi{10.3847/1538-4365/acfd29}

\bibitem[{{Cao} {et~al.}(2024{\natexlab{b}}){Cao}, {Aharonian}, {Axikegu}, {Bai}, {Bao}, {Bastieri}, {Bi}, {Bi}, {Bian}, {Bukevich}, {Cao}, {Cao}, {Cao}, {Chang}, {Chang}, {Chen}, {Chen}, {Chen}, {Chen}, {Chen}, {Chen}, {Chen}, {Chen}, {Chen}, {Chen}, {Chen}, {Chen}, {Chen}, {Chen}, {Cheng}, {Cheng}, {Cui}, {Cui}, {Cui}, {Cui}, {Dai}, {Dai}, {Dai}, {Danzengluobu}, {Dong}, {Duan}, {Fan}, {Fan}, {Fang}, {Fang}, {Fang}, {Feng}, {Feng}, {Feng}, {Feng}, {Feng}, {Feng}, {Feng}, {Gabici}, {Gao}, {Gao}, {Gao}, {Gao}, {Gao}, {Ge}, {Geng}, {Giacinti}, {Gong}, {Gou}, {Gu}, {Guo}, {Guo}, {Guo}, {Guo}, {Han}, {Hasan}, {He}, {He}, {He}, {He}, {Hor}, {Hou}, {Hou}, {Hou}, {Hu}, {Hu}, {Hu}, {Huang}, {Huang}, {Huang}, {Huang}, {Huang}, {Huang}, {Ji}, {Jia}, {Jia}, {Jiang}, {Jiang}, {Jiang}, {Jin}, {Kang}, {Karpikov}, {Kuleshov}, {Kurinov}, {Li}, {Li}, {Li}, {Li}, {Li}, {Li}, {Li}, {Li}, {Li}, {Li}, {Li}, {Li}, {Li}, {Li}, {Li}, {Li}, {Li}, {Li}, {Li}, {Liang}, {Liang}, {Lin}, {Liu}, {Liu}, {Liu}, {Liu}, {Liu}, {Liu}, {Liu},
  {Liu}, {Liu}, {Liu}, {Liu}, {Liu}, {Liu}, {Liu}, {Luo}, {Luo}, {Lv}, {Ma}, {Ma}, {Ma}, {Mao}, {Min}, {Mitthumsiri}, {Mu}, {Nan}, {Neronov}, {Ou}, {Pattarakijwanich}, {Pei}, {Qi}, {Qi}, {Qiao}, {Qin}, {Raza}, {Ruffolo}, {S{\'a}iz}, {Saeed}, {Semikoz}, {Shao}, {Shchegolev}, {Sheng}, {Shu}, {Song}, {Stenkin}, {Stepanov}, {Su}, {Sun}, {Sun}, {Sun}, {Sun}, {Takata}, {Tam}, {Tang}, {Tang}, {Tang}, {Tian}, {Wang}, {Wang}, {Wang}, {Wang}, {Wang}, {Wang}, {Wang}, {Wang}, {Wang}, {Wang}, {Wang}, {Wang}, {Wang}, {Wang}, {Wang}, {Wang}, {Wang}, {Wang}, {Wang}, {Wang}, {Wang}, {Wang}, {Wei}, {Wei}, {Wei}, {Wen}, {Wu}, {Wu}, {Wu}, {Wu}, {Wu}, {Wu}, {Xi}, {Xia}, {Xiang}, {Xiao}, {Xiao}, {Xin}, {Xing}, {Xiong}, {Xiong}, {Xu}, {Xu}, {Xu}, {Xu}, {Xue}, {Yan}, {Yan}, {Yan}, {Yang}, {Yang}, {Yang}, {Yang}, {Yang}, {Yang}, {Yang}, {Yang}, {Yao}, {Yao}, {Yin}, {Yin}, {You}, {You}, {Yu}, {Yuan}, {Yue}, {Zeng}, {Zeng}, {Zeng}, {Zha}, {Zhang}, {Zhang}, {Zhang}, {Zhang}, {Zhang}, {Zhang}, {Zhang}, {Zhang}, {Zhang}, {Zhang}, {Zhang},
  {Zhang}, {Zhang}, {Zhang}, {Zhang}, {Zhang}, {Zhang}, {Zhang}, {Zhao}, {Zhao}, {Zhao}, {Zhao}, {Zhao}, {Zhao}, {Zheng}, {Zhong}, {Zhou}, {Zhou}, {Zhou}, {Zhou}, {Zhou}, {Zhou}, {Zhou}, {Zhou}, {Zhu}, {Zhu}, {Zhu}, {Zhu}, {Zhu}, {Zou}, {Zuo}, \& {Lhaaso Collaboration}}]{2024ApJ...971L..45C}
{Cao}, Z., {Aharonian}, F., {Axikegu}, {et~al.} 2024{\natexlab{b}}, \apjl, 971, L45, \dodoi{10.3847/2041-8213/ad5e6d}

\bibitem[{{de Gasperin} {et~al.}(2011){de Gasperin}, {Merloni}, {Sell}, {Best}, {Heinz}, \& {Kauffmann}}]{2011MNRAS.415.2910D}
{de Gasperin}, F., {Merloni}, A., {Sell}, P., {et~al.} 2011, \mnras, 415, 2910, \dodoi{10.1111/j.1365-2966.2011.18904.x}

\bibitem[{{Diaz} {et~al.}(2023){Diaz}, {Hern{\`a}ndez-Garc{\'\i}a}, {Ar{\'e}valo}, {L{\'o}pez-Navas}, {Ricci}, {Koss}, {Gonzalez-Martin}, {Balokovi{\'c}}, {Osorio-Clavijo}, {Garc{\'\i}a}, \& {Malizia}}]{2023A&A...669A.114D}
{Diaz}, Y., {Hern{\`a}ndez-Garc{\'\i}a}, L., {Ar{\'e}valo}, P., {et~al.} 2023, \aap, 669, A114, \dodoi{10.1051/0004-6361/202244678}

\bibitem[{{Donato} {et~al.}(2004){Donato}, {Sambruna}, \& {Gliozzi}}]{2004ApJ...617..915D}
{Donato}, D., {Sambruna}, R.~M., \& {Gliozzi}, M. 2004, \apj, 617, 915, \dodoi{10.1086/425575}

\bibitem[{Efron(1979)}]{MR0515681}
Efron, B. 1979, Ann. Statist., 7, 1.
\newblock \url{http://links.jstor.org/sici?sici=0090-5364(197901)7:1<1:BMALAT>2.0.CO;2-6&origin=MSN}

\bibitem[{{Evans} {et~al.}(2006){Evans}, {Worrall}, {Hardcastle}, {Kraft}, \& {Birkinshaw}}]{2006ApJ...642...96E}
{Evans}, D.~A., {Worrall}, D.~M., {Hardcastle}, M.~J., {Kraft}, R.~P., \& {Birkinshaw}, M. 2006, \apj, 642, 96, \dodoi{10.1086/500658}

\bibitem[{{Fabian} {et~al.}(2000){Fabian}, {Iwasawa}, {Reynolds}, \& {Young}}]{2000PASP..112.1145F}
{Fabian}, A.~C., {Iwasawa}, K., {Reynolds}, C.~S., \& {Young}, A.~J. 2000, \pasp, 112, 1145, \dodoi{10.1086/316610}

\bibitem[{{Fanali} {et~al.}(2013){Fanali}, {Caccianiga}, {Severgnini}, {Della Ceca}, {Marchese}, {Carrera}, {Corral}, \& {Mateos}}]{2013MNRAS.433..648F}
{Fanali}, R., {Caccianiga}, A., {Severgnini}, P., {et~al.} 2013, \mnras, 433, 648, \dodoi{10.1093/mnras/stt757}

\bibitem[{{Fanaroff} \& {Riley}(1974)}]{1974MNRAS.167P..31F}
{Fanaroff}, B.~L., \& {Riley}, J.~M. 1974, \mnras, 167, 31P, \dodoi{10.1093/mnras/167.1.31P}

\bibitem[{{Fanti} {et~al.}(1995){Fanti}, {Fanti}, {Dallacasa}, {Schilizzi}, {Spencer}, \& {Stanghellini}}]{1995A&A...302..317F}
{Fanti}, C., {Fanti}, R., {Dallacasa}, D., {et~al.} 1995, \aap, 302, 317

\bibitem[{{Fanti} {et~al.}(2011){Fanti}, {Fanti}, {Zanichelli}, {Dallacasa}, \& {Stanghellini}}]{2011A&A...528A.110F}
{Fanti}, C., {Fanti}, R., {Zanichelli}, A., {Dallacasa}, D., \& {Stanghellini}, C. 2011, \aap, 528, A110, \dodoi{10.1051/0004-6361/201015379}

\bibitem[{{Gan} {et~al.}(2024){Gan}, {Zhang}, {Yang}, {Gu}, \& {Zhang}}]{2024RAA....24b5018G}
{Gan}, Y.-Y., {Zhang}, H.-M., {Yang}, X., {Gu}, Y., \& {Zhang}, J. 2024, Research in Astronomy and Astrophysics, 24, 025018, \dodoi{10.1088/1674-4527/ad1c78}

\bibitem[{{Gan} {et~al.}(2021){Gan}, {Zhang}, {Zhang}, {Yang}, {Yi}, {Liang}, \& {Liang}}]{2021RAA....21..201G}
{Gan}, Y.-Y., {Zhang}, H.-M., {Zhang}, J., {et~al.} 2021, Research in Astronomy and Astrophysics, 21, 201, \dodoi{10.1088/1674-4527/21/8/201}

\bibitem[{{Gan} {et~al.}(2022){Gan}, {Zhang}, {Yao}, {Zhang}, {Liang}, \& {Liang}}]{2022ApJ...939...78G}
{Gan}, Y.-Y., {Zhang}, J., {Yao}, S., {et~al.} 2022, \apj, 939, 78, \dodoi{10.3847/1538-4357/ac9589}

\bibitem[{{Gandhi} {et~al.}(2017){Gandhi}, {Annuar}, {Lansbury}, {Stern}, {Alexander}, {Bauer}, {Bianchi}, {Boggs}, {Boorman}, {Brandt}, {Brightman}, {Christensen}, {Comastri}, {Craig}, {Del Moro}, {Elvis}, {Guainazzi}, {Hailey}, {Harrison}, {Koss}, {Lamperti}, {Malaguti}, {Masini}, {Matt}, {Puccetti}, {Ricci}, {Rivers}, {Walton}, \& {Zhang}}]{2017MNRAS.467.4606G}
{Gandhi}, P., {Annuar}, A., {Lansbury}, G.~B., {et~al.} 2017, \mnras, 467, 4606, \dodoi{10.1093/mnras/stx357}

\bibitem[{{Giroletti} {et~al.}(2005){Giroletti}, {Taylor}, \& {Giovannini}}]{2005ApJ...622..178G}
{Giroletti}, M., {Taylor}, G.~B., \& {Giovannini}, G. 2005, \apj, 622, 178, \dodoi{10.1086/427898}

\bibitem[{{Goulding} {et~al.}(2016){Goulding}, {Greene}, {Ma}, {Veale}, {Bogdan}, {Nyland}, {Blakeslee}, {McConnell}, \& {Thomas}}]{2016ApJ...826..167G}
{Goulding}, A.~D., {Greene}, J.~E., {Ma}, C.-P., {et~al.} 2016, \apj, 826, 167, \dodoi{10.3847/0004-637X/826/2/167}

\bibitem[{{Green} {et~al.}(2009){Green}, {Aldcroft}, {Richards}, {Barkhouse}, {Constantin}, {Haggard}, {Karovska}, {Kim}, {Kim}, {Vikhlinin}, {Anderson}, {Mossman}, {Kashyap}, {Myers}, {Silverman}, {Wilkes}, \& {Tananbaum}}]{2009ApJ...690..644G}
{Green}, P.~J., {Aldcroft}, T.~L., {Richards}, G.~T., {et~al.} 2009, \apj, 690, 644, \dodoi{10.1088/0004-637X/690/1/644}

\bibitem[{{Gu} \& {Cao}(2009)}]{2009MNRAS.399..349G}
{Gu}, M., \& {Cao}, X. 2009, \mnras, 399, 349, \dodoi{10.1111/j.1365-2966.2009.15277.x}

\bibitem[{{Guainazzi} {et~al.}(2006){Guainazzi}, {Siemiginowska}, {Stanghellini}, {Grandi}, {Piconcelli}, \& {Azubike Ugwoke}}]{2006A&A...446...87G}
{Guainazzi}, M., {Siemiginowska}, A., {Stanghellini}, C., {et~al.} 2006, \aap, 446, 87, \dodoi{10.1051/0004-6361:20053374}

\bibitem[{{Hardcastle} {et~al.}(2006){Hardcastle}, {Evans}, \& {Croston}}]{2006MNRAS.370.1893H}
{Hardcastle}, M.~J., {Evans}, D.~A., \& {Croston}, J.~H. 2006, \mnras, 370, 1893, \dodoi{10.1111/j.1365-2966.2006.10615.x}

\bibitem[{{Hardcastle} {et~al.}(2009){Hardcastle}, {Evans}, \& {Croston}}]{2009MNRAS.396.1929H}
---. 2009, \mnras, 396, 1929, \dodoi{10.1111/j.1365-2966.2009.14887.x}

\bibitem[{{Hardcastle} \& {Worrall}(2000)}]{2000MNRAS.314..359H}
{Hardcastle}, M.~J., \& {Worrall}, D.~M. 2000, \mnras, 314, 359, \dodoi{10.1046/j.1365-8711.2000.03393.x}

\bibitem[{{Heckman} \& {Best}(2014)}]{2014ARA&A..52..589H}
{Heckman}, T.~M., \& {Best}, P.~N. 2014, \araa, 52, 589, \dodoi{10.1146/annurev-astro-081913-035722}

\bibitem[{{Heinz} {et~al.}(1998){Heinz}, {Reynolds}, \& {Begelman}}]{1998ApJ...501..126H}
{Heinz}, S., {Reynolds}, C.~S., \& {Begelman}, M.~C. 1998, \apj, 501, 126, \dodoi{10.1086/305807}

\bibitem[{{HI4PI Collaboration} {et~al.}(2016){HI4PI Collaboration}, {Ben Bekhti}, {Fl{\"o}er}, {Keller}, {Kerp}, {Lenz}, {Winkel}, {Bailin}, {Calabretta}, {Dedes}, {Ford}, {Gibson}, {Haud}, {Janowiecki}, {Kalberla}, {Lockman}, {McClure-Griffiths}, {Murphy}, {Nakanishi}, {Pisano}, \& {Staveley-Smith}}]{2016A&A...594A.116H}
{HI4PI Collaboration}, {Ben Bekhti}, N., {Fl{\"o}er}, L., {et~al.} 2016, \aap, 594, A116, \dodoi{10.1051/0004-6361/201629178}

\bibitem[{{Hine} \& {Longair}(1979)}]{1979MNRAS.188..111H}
{Hine}, R.~G., \& {Longair}, M.~S. 1979, \mnras, 188, 111, \dodoi{10.1093/mnras/188.1.111}

\bibitem[{{Iwasawa} {et~al.}(2012){Iwasawa}, {Mainieri}, {Brusa}, {Comastri}, {Gilli}, {Vignali}, {Hasinger}, {Sanders}, {Cappelluti}, {Impey}, {Koekemoer}, {Lanzuisi}, {Lusso}, {Merloni}, {Salvato}, {Taniguchi}, \& {Trump}}]{2012A&A...537A..86I}
{Iwasawa}, K., {Mainieri}, V., {Brusa}, M., {et~al.} 2012, \aap, 537, A86, \dodoi{10.1051/0004-6361/201118203}

\bibitem[{{Kellermann} {et~al.}(1989){Kellermann}, {Sramek}, {Schmidt}, {Shaffer}, \& {Green}}]{1989AJ.....98.1195K}
{Kellermann}, K.~I., {Sramek}, R., {Schmidt}, M., {Shaffer}, D.~B., \& {Green}, R. 1989, \aj, 98, 1195, \dodoi{10.1086/115207}

\bibitem[{{Kelly} {et~al.}(2007){Kelly}, {Bechtold}, {Siemiginowska}, {Aldcroft}, \& {Sobolewska}}]{2007ApJ...657..116K}
{Kelly}, B.~C., {Bechtold}, J., {Siemiginowska}, A., {Aldcroft}, T., \& {Sobolewska}, M. 2007, \apj, 657, 116, \dodoi{10.1086/510876}

\bibitem[{{Kiehlmann} {et~al.}(2024){Kiehlmann}, {Lister}, {Readhead}, {Liodakis}, {O'Neill}, {Pearson}, {Sheldahl}, {Siemiginowska}, {Tassis}, {Taylor}, \& {Wilkinson}}]{2024ApJ...961..240K}
{Kiehlmann}, S., {Lister}, M.~L., {Readhead}, A.~C.~S., {et~al.} 2024, \apj, 961, 240, \dodoi{10.3847/1538-4357/ad0c56}

\bibitem[{{Kr{\'o}l} {et~al.}(2024){Kr{\'o}l}, {Sobolewska}, {Stawarz}, {Siemiginowska}, {Migliori}, {Principe}, \& {Gurwell}}]{2024ApJ...966..201K}
{Kr{\'o}l}, D.~{\L}., {Sobolewska}, M., {Stawarz}, {\L}., {et~al.} 2024, \apj, 966, 201, \dodoi{10.3847/1538-4357/ad3632}

\bibitem[{{Kunert-Bajraszewska} {et~al.}(2014){Kunert-Bajraszewska}, {Labiano}, {Siemiginowska}, \& {Guainazzi}}]{2014MNRAS.437.3063K}
{Kunert-Bajraszewska}, M., {Labiano}, A., {Siemiginowska}, A., \& {Guainazzi}, M. 2014, \mnras, 437, 3063, \dodoi{10.1093/mnras/stt1978}

\bibitem[{{Laing} {et~al.}(1994){Laing}, {Jenkins}, {Wall}, \& {Unger}}]{1994ASPC...54..201L}
{Laing}, R.~A., {Jenkins}, C.~R., {Wall}, J.~V., \& {Unger}, S.~W. 1994, in Astronomical Society of the Pacific Conference Series, Vol.~54, The Physics of Active Galaxies, ed. G.~V. {Bicknell}, M.~A. {Dopita}, \& P.~J. {Quinn}, 201

\bibitem[{{Li}(2019)}]{2019MNRAS.490.3793L}
{Li}, S.-L. 2019, \mnras, 490, 3793, \dodoi{10.1093/mnras/stz2864}

\bibitem[{{Lian} {et~al.}(2024){Lian}, {Li}, {Hu}, {Gan}, {Wu}, {Zhang}, \& {Zhang}}]{2024arXiv240500347L}
{Lian}, J.-S., {Li}, J.-X., {Hu}, X.-K., {et~al.} 2024, arXiv e-prints, arXiv:2405.00347, \dodoi{10.48550/arXiv.2405.00347}

\bibitem[{{Liao} \& {Gu}(2020)}]{2020MNRAS.491...92L}
{Liao}, M., \& {Gu}, M. 2020, \mnras, 491, 92, \dodoi{10.1093/mnras/stz2981}

\bibitem[{{Liao} {et~al.}(2020){Liao}, {Gu}, {Zhou}, \& {Chen}}]{2020MNRAS.497..482L}
{Liao}, M., {Gu}, M., {Zhou}, M., \& {Chen}, L. 2020, \mnras, 497, 482, \dodoi{10.1093/mnras/staa1559}

\bibitem[{{Lister} {et~al.}(2020){Lister}, {Homan}, {Kovalev}, {Mandal}, {Pushkarev}, \& {Siemiginowska}}]{2020ApJ...899..141L}
{Lister}, M.~L., {Homan}, D.~C., {Kovalev}, Y.~Y., {et~al.} 2020, \apj, 899, 141, \dodoi{10.3847/1538-4357/aba18d}

\bibitem[{{Migliori} {et~al.}(2016){Migliori}, {Siemiginowska}, {Sobolewska}, {Loh}, {Corbel}, {Ostorero}, \& {Stawarz}}]{2016ApJ...821L..31M}
{Migliori}, G., {Siemiginowska}, A., {Sobolewska}, M., {et~al.} 2016, \apjl, 821, L31, \dodoi{10.3847/2041-8205/821/2/L31}

\bibitem[{{Murgia}(2003)}]{2003PASA...20...19M}
{Murgia}, M. 2003, \pasa, 20, 19, \dodoi{10.1071/AS02033}

\bibitem[{{Nandra} {et~al.}(2007){Nandra}, {O'Neill}, {George}, \& {Reeves}}]{2007MNRAS.382..194N}
{Nandra}, K., {O'Neill}, P.~M., {George}, I.~M., \& {Reeves}, J.~N. 2007, \mnras, 382, 194, \dodoi{10.1111/j.1365-2966.2007.12331.x}

\bibitem[{{O'Dea} {et~al.}(2000){O'Dea}, {De Vries}, {Worrall}, {Baum}, \& {Koekemoer}}]{2000AJ....119..478O}
{O'Dea}, C.~P., {De Vries}, W.~H., {Worrall}, D.~M., {Baum}, S.~A., \& {Koekemoer}, A. 2000, \aj, 119, 478, \dodoi{10.1086/301209}

\bibitem[{{Ostorero} {et~al.}(2010){Ostorero}, {Moderski}, {Stawarz}, {Diaferio}, {Kowalska}, {Cheung}, {Kataoka}, {Begelman}, \& {Wagner}}]{2010ApJ...715.1071O}
{Ostorero}, L., {Moderski}, R., {Stawarz}, {\L}., {et~al.} 2010, \apj, 715, 1071, \dodoi{10.1088/0004-637X/715/2/1071}

\bibitem[{{O'Sullivan} {et~al.}(2003){O'Sullivan}, {Ponman}, \& {Collins}}]{2003MNRAS.340.1375O}
{O'Sullivan}, E., {Ponman}, T.~J., \& {Collins}, R.~S. 2003, \mnras, 340, 1375, \dodoi{10.1046/j.1365-8711.2003.06396.x}

\bibitem[{{Owsianik} \& {Conway}(1998)}]{1998A&A...337...69O}
{Owsianik}, I., \& {Conway}, J.~E. 1998, \aap, 337, 69, \dodoi{10.48550/arXiv.astro-ph/9712062}

\bibitem[{{Phillips} \& {Mutel}(1980)}]{1980ApJ...236...89P}
{Phillips}, R.~B., \& {Mutel}, R.~L. 1980, \apj, 236, 89, \dodoi{10.1086/157722}

\bibitem[{{Principe} {et~al.}(2020){Principe}, {Migliori}, {Johnson}, {D'Ammando}, {Giroletti}, {Orienti}, {Stanghellini}, {Taylor}, {Torresi}, \& {Cheung}}]{2020A&A...635A.185P}
{Principe}, G., {Migliori}, G., {Johnson}, T.~J., {et~al.} 2020, \aap, 635, A185, \dodoi{10.1051/0004-6361/201937049}

\bibitem[{{Readhead} {et~al.}(1996{\natexlab{a}}){Readhead}, {Taylor}, {Pearson}, \& {Wilkinson}}]{1996ApJ...460..634R}
{Readhead}, A.~C.~S., {Taylor}, G.~B., {Pearson}, T.~J., \& {Wilkinson}, P.~N. 1996{\natexlab{a}}, \apj, 460, 634, \dodoi{10.1086/176997}

\bibitem[{{Readhead} {et~al.}(1996{\natexlab{b}}){Readhead}, {Taylor}, {Xu}, {Pearson}, {Wilkinson}, \& {Polatidis}}]{1996ApJ...460..612R}
{Readhead}, A.~C.~S., {Taylor}, G.~B., {Xu}, W., {et~al.} 1996{\natexlab{b}}, \apj, 460, 612, \dodoi{10.1086/176996}

\bibitem[{{Readhead} {et~al.}(2021){Readhead}, {Ravi}, {Liodakis}, {Lister}, {Singh}, {Aller}, {Blandford}, {Browne}, {Gorjian}, {Grainge}, {Gurwell}, {Hodges}, {Hovatta}, {Kiehlmann}, {L{\"a}hteenm{\"a}ki}, {Mcaloone}, {Max-Moerbeck}, {Pavlidou}, {Pearson}, {Peirson}, {Perlman}, {Reeves}, {Soifer}, {Taylor}, {Tornikoski}, {Vedantham}, {Werner}, {Wilkinson}, \& {Zensus}}]{2021ApJ...907...61R}
{Readhead}, A.~C.~S., {Ravi}, V., {Liodakis}, I., {et~al.} 2021, \apj, 907, 61, \dodoi{10.3847/1538-4357/abd08c}

\bibitem[{{Reeves} \& {Turner}(2000)}]{2000MNRAS.316..234R}
{Reeves}, J.~N., \& {Turner}, M.~J.~L. 2000, \mnras, 316, 234, \dodoi{10.1046/j.1365-8711.2000.03510.x}

\bibitem[{{Ricci} {et~al.}(2014){Ricci}, {Ueda}, {Paltani}, {Ichikawa}, {Gandhi}, \& {Awaki}}]{2014MNRAS.441.3622R}
{Ricci}, C., {Ueda}, Y., {Paltani}, S., {et~al.} 2014, \mnras, 441, 3622, \dodoi{10.1093/mnras/stu735}

\bibitem[{{Risaliti} {et~al.}(2009){Risaliti}, {Young}, \& {Elvis}}]{2009ApJ...700L...6R}
{Risaliti}, G., {Young}, M., \& {Elvis}, M. 2009, \apjl, 700, L6, \dodoi{10.1088/0004-637X/700/1/L6}

\bibitem[{{She} {et~al.}(2018){She}, {Ho}, {Feng}, \& {Cui}}]{2018ApJ...859..152S}
{She}, R., {Ho}, L.~C., {Feng}, H., \& {Cui}, C. 2018, \apj, 859, 152, \dodoi{10.3847/1538-4357/aabfe7}

\bibitem[{{Shemmer} {et~al.}(2006){Shemmer}, {Brandt}, {Netzer}, {Maiolino}, \& {Kaspi}}]{2006ApJ...646L..29S}
{Shemmer}, O., {Brandt}, W.~N., {Netzer}, H., {Maiolino}, R., \& {Kaspi}, S. 2006, \apjl, 646, L29, \dodoi{10.1086/506911}

\bibitem[{{Shemmer} {et~al.}(2008){Shemmer}, {Brandt}, {Netzer}, {Maiolino}, \& {Kaspi}}]{2008ApJ...682...81S}
---. 2008, \apj, 682, 81, \dodoi{10.1086/588776}

\bibitem[{{Siemiginowska} {et~al.}(2008){Siemiginowska}, {LaMassa}, {Aldcroft}, {Bechtold}, \& {Elvis}}]{2008ApJ...684..811S}
{Siemiginowska}, A., {LaMassa}, S., {Aldcroft}, T.~L., {Bechtold}, J., \& {Elvis}, M. 2008, \apj, 684, 811, \dodoi{10.1086/589437}

\bibitem[{{Siemiginowska} {et~al.}(2016){Siemiginowska}, {Sobolewska}, {Migliori}, {Guainazzi}, {Hardcastle}, {Ostorero}, \& {Stawarz}}]{2016ApJ...823...57S}
{Siemiginowska}, A., {Sobolewska}, M., {Migliori}, G., {et~al.} 2016, \apj, 823, 57, \dodoi{10.3847/0004-637X/823/1/57}

\bibitem[{{Sobolewska} {et~al.}(2019{\natexlab{a}}){Sobolewska}, {Siemiginowska}, {Guainazzi}, {Hardcastle}, {Migliori}, {Ostorero}, \& {Stawarz}}]{2019ApJ...884..166S}
{Sobolewska}, M., {Siemiginowska}, A., {Guainazzi}, M., {et~al.} 2019{\natexlab{a}}, \apj, 884, 166, \dodoi{10.3847/1538-4357/ab3ec3}

\bibitem[{{Sobolewska} {et~al.}(2019{\natexlab{b}}){Sobolewska}, {Siemiginowska}, {Guainazzi}, {Hardcastle}, {Migliori}, {Ostorero}, \& {Stawarz}}]{2019ApJ...871...71S}
---. 2019{\natexlab{b}}, \apj, 871, 71, \dodoi{10.3847/1538-4357/aaee78}

\bibitem[{{Sobolewska} {et~al.}(2023){Sobolewska}, {Siemiginowska}, {Migliori}, {Ostorero}, {Stawarz}, \& {Guainazzi}}]{2023ApJ...948...81S}
{Sobolewska}, M., {Siemiginowska}, A., {Migliori}, G., {et~al.} 2023, \apj, 948, 81, \dodoi{10.3847/1538-4357/acbb6c}

\bibitem[{{Stawarz} {et~al.}(2008){Stawarz}, {Ostorero}, {Begelman}, {Moderski}, {Kataoka}, \& {Wagner}}]{2008ApJ...680..911S}
{Stawarz}, {\L}., {Ostorero}, L., {Begelman}, M.~C., {et~al.} 2008, \apj, 680, 911, \dodoi{10.1086/587781}

\bibitem[{{Taylor} {et~al.}(2000){Taylor}, {Marr}, {Pearson}, \& {Readhead}}]{2000ApJ...541..112T}
{Taylor}, G.~B., {Marr}, J.~M., {Pearson}, T.~J., \& {Readhead}, A.~C.~S. 2000, \apj, 541, 112, \dodoi{10.1086/309428}

\bibitem[{{Tengstrand} {et~al.}(2009){Tengstrand}, {Guainazzi}, {Siemiginowska}, {Fonseca Bonilla}, {Labiano}, {Worrall}, {Grandi}, \& {Piconcelli}}]{2009A&A...501...89T}
{Tengstrand}, O., {Guainazzi}, M., {Siemiginowska}, A., {et~al.} 2009, \aap, 501, 89, \dodoi{10.1051/0004-6361/200811284}

\bibitem[{{Turner} \& {Miller}(2009)}]{2009A&ARv..17...47T}
{Turner}, T.~J., \& {Miller}, L. 2009, \aapr, 17, 47, \dodoi{10.1007/s00159-009-0017-1}

\bibitem[{{Urry} \& {Padovani}(1995)}]{1995PASP..107..803U}
{Urry}, C.~M., \& {Padovani}, P. 1995, \pasp, 107, 803, \dodoi{10.1086/133630}

\bibitem[{{van Breugel} {et~al.}(1984){van Breugel}, {Miley}, \& {Heckman}}]{1984AJ.....89....5V}
{van Breugel}, W., {Miley}, G., \& {Heckman}, T. 1984, \aj, 89, 5, \dodoi{10.1086/113480}

\bibitem[{{Vink} {et~al.}(2006){Vink}, {Snellen}, {Mack}, \& {Schilizzi}}]{2006MNRAS.367..928V}
{Vink}, J., {Snellen}, I., {Mack}, K.-H., \& {Schilizzi}, R. 2006, \mnras, 367, 928, \dodoi{10.1111/j.1365-2966.2006.10036.x}

\bibitem[{{Wang} {et~al.}(2004){Wang}, {Watarai}, \& {Mineshige}}]{2004ApJ...607L.107W}
{Wang}, J.-M., {Watarai}, K.-Y., \& {Mineshige}, S. 2004, \apjl, 607, L107, \dodoi{10.1086/421906}

\bibitem[{{Wilkinson} {et~al.}(1994){Wilkinson}, {Polatidis}, {Readhead}, {Xu}, \& {Pearson}}]{1994ApJ...432L..87W}
{Wilkinson}, P.~N., {Polatidis}, A.~G., {Readhead}, A.~C.~S., {Xu}, W., \& {Pearson}, T.~J. 1994, \apjl, 432, L87, \dodoi{10.1086/187518}

\bibitem[{{W{\'o}jtowicz} {et~al.}(2020){W{\'o}jtowicz}, {Stawarz}, {Cheung}, {Ostorero}, {Kosmaczewski}, \& {Siemiginowska}}]{2020ApJ...892..116W}
{W{\'o}jtowicz}, A., {Stawarz}, {\l}., {Cheung}, C.~C., {et~al.} 2020, \apj, 892, 116, \dodoi{10.3847/1538-4357/ab7930}

\bibitem[{{Woo} \& {Urry}(2002)}]{2002ApJ...579..530W}
{Woo}, J.-H., \& {Urry}, C.~M. 2002, \apj, 579, 530, \dodoi{10.1086/342878}

\bibitem[{{Yang} {et~al.}(2015){Yang}, {Xie}, {Yuan}, {Zdziarski}, {Gierli{\'n}ski}, {Ho}, \& {Yu}}]{2015MNRAS.447.1692Y}
{Yang}, Q.-X., {Xie}, F.-G., {Yuan}, F., {et~al.} 2015, \mnras, 447, 1692, \dodoi{10.1093/mnras/stu2571}

\bibitem[{{Younes} {et~al.}(2011){Younes}, {Porquet}, {Sabra}, \& {Reeves}}]{2011A&A...530A.149Y}
{Younes}, G., {Porquet}, D., {Sabra}, B., \& {Reeves}, J.~N. 2011, \aap, 530, A149, \dodoi{10.1051/0004-6361/201116806}

\bibitem[{{Yuan} \& {Narayan}(2014)}]{2014ARA&A..52..529Y}
{Yuan}, F., \& {Narayan}, R. 2014, \araa, 52, 529, \dodoi{10.1146/annurev-astro-082812-141003}

\end{thebibliography}
\clearpage
\appendix

\section{4 Obscured CSOs} \label{append_4CSO}

\emph{OQ 208}. By combining the observational data from \xmm, \chandra, and \emph{NuSTAR}, \cite{2019ApJ...884..166S} constructed a broadband X-ray spectrum. They obtained a high equivalent column density of $10^{23-24}$ cm$^{-2}$ with a photon spectral index of $\Gamma_{\rm X}=1.45_{-0.01}^{+0.11}$ and the 2-10 keV luminosity of $L_{\rm 2-10~keV}\sim4.5 \times 10^{42}$ erg s$^{-1}$, and they proposed that the presence of cold matter obstructs the radio jet in OQ 208.

\emph{JVAS J1511+0518}. By combining the observational data from \xmm\, and \emph{NuSTAR}, \cite{2023ApJ...948...81S} constructed a broadband X-ray spectrum. They found that a toroidal reprocessor model can reproduce the spectrum well, the equivalent column density is $\sim 10^{24}$ cm$^{-2}$ with a photon spectral index of $\Gamma_{\rm X}=1.70_{-0.05}^{+0.15}$ and the 2-10 keV luminosity of $L_{\rm 2-10~keV}\sim3.8 \times 10^{42}$ erg s$^{-1}$.

\emph{S4 2021+61}. By combining the observational data from \xmm\, and \emph{NuSTAR}, \cite{2023ApJ...948...81S} also constructed its broadband X-ray spectrum, which is also fitted with a toroidal reprocessor model. The derived equivalent column density is $\sim 10^{24}$ cm$^{-2}$ with a photon spectral index of $\Gamma_{\rm X}=1.45_{-0.05}^{+0.09}$ and the 2-10 keV luminosity of $L_{\rm 2-10~keV}\sim1.1 \times 10^{44}$ erg s$^{-1}$.

\emph{NGC 7674}. By combining the observational data from \emph{NuSTAR}, \emph{Swift-XRT}, and \emph{Suzaku}, \cite{2017MNRAS.467.4606G} assumed geometry of the nuclear obscurer/reﬂector to carry out the X-ray spectrum modeling in the 0.5-78 keV band. They obtained the column density of $3.4 \times 10^{24}$ cm$^{-2}$ with a photon spectral index of $\Gamma_{\rm X}=2.07_{-0.11}^{+0.15}$ and the 2-10 keV luminosity of $L_{\rm 2-10~keV}\sim3-5\times 10^{43}$ erg s$^{-1}$.

The information regarding the four obscured CSOs is also included in Table \ref{tab-4CSOs}.

\section{Analysis of X-ray Data for Individual CSOs} \label{append_indiv}

In this study, we acquired 18 valid X-ray spectral fits from 12 \chandra\, observations and 14 valid X-ray spectral fits from 12 \xmm\, observations, which covered 17 out of 22 CSOs. Representative spectral fit results for each source are illustrated in Figures \ref{spect-Chandra} and \ref{spect-XMM}. Detailed observations and data analysis information on 22 CSOs are provided below.

\emph{B2 0026+34}. The X-ray spectrum obtained with the \xmm\, observation on 2004 January 08 is well described by an absorbed power-law model, yielding $N_{\rm H}^{\rm int}=(8.5_{-3.1}^{+3.6}) \times 10^{21}$ cm$^{-2}$ and $\Gamma_{\rm X}=1.46_{-0.20}^{+0.21}$. The corrected flux in the 2--10 keV energy range, denoted as $F_{\rm 2-10 keV}$, is $(2.44_{-0.35}^{+0.32}) \times 10^{-13}$ erg cm$^{-2}$ s$^{-1}$. These findings are consistent with those reported by \cite{2006A&A...446...87G}, considering the associated uncertainties.

\emph{0108+388}. \xmm\, observed it on 2004 January 09. The extracted X-ray spectrum is unsuitable for spectral fitting due to low signal-to-noise ratio.

\emph{B2 0116+31}. It is only observed once by \chandra\, on 2010 November 04. However, the recorded net photon counts are insufficient to generate a valid X-ray spectrum.

\emph{TXS 0128+554}. It is a $\gamma$-ray emitting CSO (\citealp{2020ApJ...899..141L, 2024RAA....24b5018G}). \chandra\, conducted a series of daily observations from 2019 March 29 to 2019 April 01, totaling four observations. We extract the X-ray spectrum from each observation and perform a joint spectral fitting using an absorbed power-law model. The joint spectral fitting yields $N_{\rm H}^{\rm int}=(3.4_{-1.9}^{+2.1}) \times 10^{21}$ cm$^{-2}$. Subsequently, we fix this $N_{\rm H}^{\rm int}$ value for the spectral fitting of other observations and derive the values of $\Gamma_{\rm X}$ and $F_{\rm 2-10~keV}$ for each observation. Our analysis reveals that TXS 0128+554 does not exhibit any discernible X-ray variability or spectral evolution on daily timescales. \cite{2020ApJ...899..141L} combined the four observations and derived the spectrum of TXS 0128+554. By employing the same model, they obtained $N_{\rm H}^{\rm int} = (7.0 \pm 0.8) \times 10^{21}$ cm$^{-2}$, $\Gamma_{\rm X} = 2.38 \pm 0.10$, and an unabsorbed flux of $(7.04_{-1.23}^{+1.45}) \times 10^{-13}$ erg cm$^{-2}$ s$^{-1}$ in the 2--10 keV band.

\emph{B3 0402+379}. The \chandra\, observations were conducted on 2011 April 04 and 2013 November 06. We first employ an absorbed power-law model to fit the joint spectra of the two observations. However, we identify the presence of an emission line at approximately 6.4 keV, necessitating the inclusion of a Gaussian component in addition to the absorbed power-law model. During spectral fitting, the width of the Gaussian line ($\sigma_\mathrm{l}$) was fixed at 0.01 keV, yielding an energy of $E_\mathrm{l} = 6.57_{-0.05}^{+0.05}$ keV for the Gaussian line. This detected emission line is likely the iron K$\alpha$ line. We derive an intrinsic absorbing column density of $N_{\rm H}^{\rm int} = (7.1_{-1.5}^{+1.7}) \times 10^{21}$ cm$^{-2}$, which was subsequently held constant during the fitting of each X-ray spectrum. Due to insufficient statistical data, no discernible emission line signal was detected in the observation on 2011 April 04. The X-ray spectrum from this observation is well-fitted by an absorbed power-law model. Based on the results of spectral fitting, it is observed that the X-ray spectrum became steeper from 2011 April 04 to 2013 November 06, with $\Gamma_{\rm X}$ changing from $2.60_{-0.46}^{+0.52}$ to $2.97_{-0.11}^{+0.12}$. However, considering the associated errors, $\Gamma_{\rm X}$ remains relatively consistent between these two observations. The values of $F_{\rm 2-10~keV}$ for both observations are very similar.

\emph{B3 0710+439}. Both \xmm\, and \chandra\, observed the object on 2004 March 22 and 2011 January 18, respectively. The combined spectra from these two observations can be accurately modeled by an absorbed power-law model. We determine the intrinsic absorbing column density to be $N_{\rm H}^{\rm int} = (4.3_{-1.0}^{+1.1}) \times 10^{21}$ cm$^{-2}$, which is held constant during the fitting of the individual spectra.  Both spectra exhibit a notably hard nature, with spectral indices of \(1.59 \pm 0.07\) for the 2004 observation and \(1.49 \pm 0.07\) for the 2011 observation, consistent within the uncertainties. The corrected flux in the 2-10 keV energy range is $(3.67_{-0.33}^{+0.31}) \times 10^{-13}$ erg cm$^{-2}$ s$^{-1}$ on 2004 March 22, and increasing to $(4.76_{-0.40}^{+0.39}) \times 10^{-13}$ erg cm$^{-2}$ s$^{-1}$ on 2011 January 18. These findings are consistent with previous reports by \cite{2006MNRAS.367..928V} and \cite{2016ApJ...823...57S}. 

\emph{JVAS J1035+5628}. The observation was conducted using \xmm\, on 2004 October 21. However, persistent high background burst activity significantly impacts the observation for most of the duration. After excluding the affected periods, we manage to extract the X-ray spectrum of the source; however, its signal-to-noise ratio is insufficient for reliable spectral fitting. \cite{2006MNRAS.367..928V} also analyzed this observation and employed high cutoff rates to exclude the time intervals with high background activity, the cutoff rates adopted in their work are considerably higher than those used in our analysis.

\emph{PKS 1117+146}. It was observed once by \xmm\, on 2007 June 13. In the obtained image, another source in close proximity to PKS 1117+146 is identified; therefore, the spectrum of PKS 1117+146 is extracted within a range of $\sim16^{\prime\prime}$ to avoid contamination. The X-ray spectrum derived from this observation can be reproduced by an absorbed power-law model with $\Gamma_{\rm X} = 1.41_{-0.58}^{+0.74}$. Only an upper limit for the intrinsic column density could be estimated, i.e., $N_{\rm H}^{\rm int}<2.4 \times 10^{21}$ cm$^{-2}$. The corrected flux in the energy range of 2-10 keV is determined as $F_{\rm 2-10~keV}=(1.6_{-0.6}^{+0.4}) \times 10^{-14}$ erg cm$^{-2}$ s$^{-1}$, consistent with previous findings reported in \cite{2009A&A...501...89T}.

\emph{NGC 3894}. It is also a $\gamma$-ray emitting CSO (\citealp{2020A&A...635A.185P, 2024RAA....24b5018G}). The object has been observed once by \chandra\, on 2009 July 20 and twice by \xmm\, on 2022 November 14. However, one of the \xmm\, observations did not yield any significant detections. Therefore, we can only utilize the X-ray spectra from two observations. Within a radius of $20^{\prime\prime}$ around the location of NGC 3894 in the \chandra\, observation image, three X-ray sources were identified. However, their impact is deemed negligible and thus not included in our analysis. Describing the two spectra using only an absorbed power-law model is challenging and requires incorporating a second component to account for the excess in the soft X-ray band. This second component is presumed to be a thermal plasma component, given that extended X-ray emission in the soft X-ray band has been observed for NGC 3894 by \chandra\, (e.g., \citealp{2021ApJ...922...84B}). Additionally, we identify an emission line signature at approximately 6.4 keV in \chandra\, spectrum, which prompt us to include a Gaussian component into the fitting process. However, \xmm\,'s spectrum  does not provide sufficient constraints for accurately determining the parameters of the Gaussian component. During spectral fitting, we fix the Gaussian line width $\sigma_\mathrm{l}$ as 0.01 keV. For the spectrum extracted from \chandra\, observation, we obtain $N_{\rm H}^{\rm int} = (1.58_{-0.88}^{+1.18}) \times 10^{22}$ cm$^{-2}$, $\Gamma = 0.95_{-0.53}^{+0.61}$, plasma temperature of $kT = 0.85_{-0.09}^{+0.11}$ keV, the Gaussian line energy $E_\mathrm{l} = 6.42_{-0.10}^{+0.11}$ keV, and intrinsic flux $F_{\rm 2-10keV} = (2.47_{-0.44}^{+0.42}) \times 10^{-13}$ erg cm$^{-2}$ s$^{-1}$. For \xmm\, observation, we obtain $N_{\rm H}^{\rm int} < 2 \times 10^{20}$ cm$^{-2}$, $\Gamma = 0.85_{-0.10}^{+0.10}$, plasma temperature of $kT = 0.36_{-0.04}^{+0.07}$ keV, and intrinsic flux $F_{\rm 2-10keV} = \rm (1.77_{-0.12}^{+0.13}) \times 10^{-13}$ erg cm$^{-2}$ s$^{-1}$. 

\emph{JVAS J1234+4753}. The X-ray observations were conducted by \chandra\, on 2002 March 23 and \xmm\, on 2004 June 17. However, the \xmm\, observation was affected by persistent high-background bursts, which precluded obtaining an effective PN spectrum. Additionally, the \chandra\, observation experienced a significant pileup effect, as detailed in Section 3.1. We attempt to combine the \chandra\, and MOS spectra for joint fitting; however, due to the inability to constrain the $N_{\rm H}^{\rm int}$ value, we fit these two spectra separately. An absorbed power-law model is employed to fit the two X-ray spectra, resulting in $N_{\rm H}^{\rm int}<4.4 \times 10^{21}$ cm$^{-2}$ and $\Gamma_{\rm X}=2.45_{-0.47}^{+0.96}$ for the \chandra\, spectrum, and $N_{\rm H}^{\rm int}<1.9 \times 10^{21}$ cm$^{-2}$ and $\Gamma_{\rm X}=2.21_{-0.23}^{+0.44}$ for the MOS spectrum. \cite{2009ApJ...690..644G} reported $N_{\rm H}^{\rm int}=(0.6_{-0.6}^{+1.2}) \times 10^{21}$ cm$^{-2}$ and $\Gamma_{\rm X}=1.80_{-0.20}^{+0.24}$ for the \chandra\, observation. The discrepancies in their fitting results can be attributed to the omission of the pileup effect in their analysis.

\emph{JVAS J1247+6723}. The observation was conducted by \xmm\, on 2005 November 29. However, due to the low signal-to-noise ratio, an X-ray spectral analysis could not be performed.

\emph{DA 344}. The \xmm\, observation was conducted on 2007 December 5. We extract the 0.5--7.5 keV spectrum and fit it with an absorbed power-law model, yielding an upper limit of $N_{\rm H}^{\rm int} < 1.2\times 10^{21}$ cm$^{-2}$ with $\Gamma_{\rm X} = 1.67_{-0.14}^{+0.20}$ and $F_{\rm 2-10~keV} =(8.1_{-1.3}^{+1.2}) \times 10^{-14}$ erg cm$^{-2}$ s$^{-1}$. Using the same model, \cite{2009A&A...501...89T} fitted the spectrum in the 2--10 keV band and obtained $N_{\rm H}^{\rm int} = (1.2_{-0.5}^{+0.6}) \times 10^{21}$ cm$^{-2}$, $\Gamma_{\rm X} = 1.74 \pm 0.2$, and an unabsorbed flux of $(7.8 \pm 1.3) \times 10^{-14}$ erg cm$^{-2}$ s$^{-1}$. The values for spectral index and flux are generally consistent with theirs; however, we only obtain an upper limit for $N_{\rm H}^{\rm int}$ possibly due to our narrower fitted energy band.

\emph{PKS B1345+125}. It was observed by \chandra\, on 2000 February 24. An absorbed power-law model inadequately describes its X-ray spectrum, as indicated by a significant excess in the soft X-ray band observed in the residuals. The quality of the spectral fit improves notably with the inclusion of a thermal plasma emission component. The fitting parameters are as follows: $\Gamma = 1.45_{-0.24}^{+0.25}$, $N_{\rm H}^{\rm int} = (3.20_{-0.56}^{+0.63}) \times 10^{22}$ cm$^{-2}$, and the plasma temperature $kT = 0.83_{-0.10}^{+0.10}$ keV. The soft X-ray thermal emission is attributed to the extended emission in PKS B1345+125, which may be related to the galaxy halo (\citealp{2008ApJ...684..811S}). Furthermore, the measured $kT$ value is consistent with those reported for PKS 1718--649 and NGC 3894, as well as with the temperatures of hot X-ray halos in several galaxy samples (e.g., \citealp{2003MNRAS.340.1375O, 2016ApJ...826..167G}).

\emph{1358+625}. The \xmm\, observed it on 2004 April 14; however, the observation was significantly impacted by persistent high background bursts. \cite{2006MNRAS.367..928V} utilized all available data of this observation without any exclusions to extract the X-ray spectrum of the source. In our analysis, we exclude the time intervals with high background activity using a reasonable count rate threshold, which consequently prevented us from generating an effective energy spectrum.

\emph{B3 1441+409}. The X-ray spectrum obtained by \xmm\, observation on 2009 January 18 can be well fitted by an absorbed power-law model. We obtain $N_{\rm H}^{\rm int} = (1.6_{-1.1}^{+1.3}) \times 10^{21}$ cm$^{-2}$, with $\Gamma_{\rm X}=1.85_{-0.31}^{+0.36}$ and $F_{\rm 2-10~keV}=(7.4_{-2.3}^{+2.0}) \times 10^{-14}$ erg cm$^{-2}$ s$^{-1}$.

\emph{4C 52.37}. It was observed by \chandra\, on 2007 June 3. Although applying a forced fit with an absorbed power-law model yields similar results to those reported in \cite{2011MNRAS.415.2910D}, the resulting fitting statistics is 14.8/5 for the C-statistic method.

\emph{CTD 93}. It has been observed once by \chandra\, on 2010 December 4 and twice by \xmm\, on 2008 January 17 and 19. However, the observations conducted by \xmm\, were hindered by persistent high-background bursts, preventing the extraction of a valid spectrum from observation (Obs-ID 0502510801). In the \chandra\, image, a secondary source is identified in the extracted background region; however, its influence can be neglected. We combine the \chandra\, and \xmm\, spectra for a joint fitting, but the $N_{\rm H}^{\rm int}$ value could not be constrained well. Therefore, we fit these spectra separately. The X-ray spectrum from \chandra\, observation is adequately described by a hard absorbed power-law model with $\Gamma_{\rm X}=1.35_{-0.24}^{+0.29}$ and $F_{\rm 2-10~keV}=(4.7_{-0.7}^{+0.6}) \times 10^{-14}$ erg cm$^{-2}$ s$^{-1}$, yielding an upper limit of $N_{\rm H}^{\rm int}<3.2 \times 10^{21}$ cm$^{-2}$. These findings are consistent with those reported in \cite{2016ApJ...823...57S}. Applying the same model to fit the MOS spectrum (Obs-ID 0502510401), we obtain $\Gamma_{\rm X}=2.24_{-0.75}^{+3.87}$ and $F_{\rm 2-10~keV}=(3.1_{-1.1}^{+0.8}) \times 10^{-14}$ erg cm$^{-2}$ s$^{-1}$, with an upper limit of $N_{\rm H}^{\rm int}<2.73 \times 10^{22}$ cm$^{-2}$.

\emph{PKS 1718--649}. It is the first identified $\gamma$-ray emitting CSO (\citealp{2016ApJ...821L..31M, 2024RAA....24b5018G}). The source was observed six times, three times by \chandra\, on 2010 November 9, 2014 June 20 and 23, and another three times by \xmm\, on 2017 March 5, 2018 March 8, and 2020 March 27. After considering an absorbed power-law model, no significant residuals are observed in the soft X-ray band of \chandra's spectra, while obvious residuals can be found in the soft X-ray band of \xmm\,’s spectra. These residuals may be attributed to the extended X-ray emission in PKS 1718--649 (e.g., \citealp{2016ApJ...823...57S}). We therefore independently merge and analyze the \chandra\, and \xmm\, spectra. For the combined \chandra\, spectrum, the model \texttt{tbabs*ztbabs(powerlaw)} in XSPEC is used for fitting. For the combined \xmm\, spectrum, the model \texttt{tbabs*(ztbabs(powerlaw)+apec)} in XSPEC is employed. Subsequently, the obtained $N_{\rm H}^{\rm int}$ values from the combined fits are fixed for each individual spectral fitting. The derived plasma temperature $kT=0.80_{-0.04}^{+0.05}$ keV is consistent with that observed in the X-ray halos of certain galaxy samples (e.g., \citealp{2003MNRAS.340.1375O, 2016ApJ...826..167G}).

\emph{PKS 1934--63}. It was observed by \chandra\, on 2010 July 8 and by \xmm\, on 2017 April 1. In the obtained images, a secondary source adjacent to PKS 1934--63 is identified. To ensure accurate measurements, we extract the source spectrum from the \xmm\, observation within a range of $\sim16^{\prime\prime}$. The spectral extraction for the \chandra\, observation remains unaffected by the presence of the second source and takes the same region range as other sources. An absorbed power-law model is used to reproduce the combined spectrum of PKS 1934--63; however, we are unable to constrain the $N_{\rm H}^{\rm int}$ value, and thus we separately fit these spectra. Both spectra obtained from \chandra\, and \xmm\, observations provide only an upper limit for $N_{\rm H}^{\rm int}$, i.e., $N_{\rm H}^{\rm int}<1.0 \times 10^{21}$ cm$^{-2}$. The photon spectral index and intrinsic 2-10 keV flux remain consistent in both observations. \cite{2016ApJ...823...57S} also analyzed the same \chandra\, observation and obtained $N_{\rm H}^{\rm int} = (8_{-6}^{+7}) \times 10^{20}$ cm$^{-2}$, $\Gamma_{\rm X} = 1.67_{-0.16}^{+0.15}$, and $F_{\rm 2-10~keV}=(1.23 \pm 0.16) \times 10^{-13}$ erg cm$^{-2}$ s$^{-1}$. The values of $\Gamma_{\rm X}$ and $F_{\rm 2-10~keV}$ are in agreement with our results. However, the source and background extracted regions in their study are significantly smaller than ours, and the background was not considered in their analysis. These differences in processing details may account for the discrepancy in the $N_{\rm H}^{\rm int}$ value.

\emph{S4 1943+54}. It was observed only once by \chandra\, on 2011 May 4. The X-ray spectrum is fitted using an absorbed power-law model. Given the low net counts, we fix the spectral index $\Gamma_{\rm X}$ at $1.7$ during fitting and obtain an upper limit of $N_{\rm H}^{\rm int}<3.26 \times 10^{22}$ cm$^{-2}$. \cite{2016ApJ...823...57S} analyzed the same observation, and they adopted the same model for fitting the X-ray spectrum, also fixing $\Gamma_{\rm X}$ at $1.7$. They reported $N_{\rm H}^{\rm int} = (1.1 \pm 0.7) \times 10^{22}$ cm$^{-2}$ and $F_{\rm 2-10~keV}=(3.7 \pm 1.4) \times 10^{-14}$ erg cm$^{-2}$ s$^{-1}$. Considering the associated uncertainties, the $F_{\rm 2-10~keV}$ value is consistent with our results. However, similar to the \chandra\, observation of PKS 1934--63, discrepancies in $N_{\rm H}^{\rm int}$ may arise from differences in data processing.

\emph{S5 1946+70}. The source was observed with \chandra\, on 2011 February 6, and with \xmm\, on 2016 October 21.  The combined spectrum from these two observations can be accurately modeled using an absorbed power-law model, yielding $N_{\rm H}^{\rm int} = (1.16_{-0.39}^{+0.51}) \times 10^{22}$ cm$^{-2}$. The $N_{\rm H}^{\rm int}$ value obtained through joint spectral fitting is fixed during the individual spectral fits. For the \chandra\, observation, the parameters are as follows: $\Gamma_{\rm X} = 1.50 \pm 0.36$ and $F_{\rm 2-10~keV}=(4.51_{-1.32}^{+1.31}) \times 10^{-13}$ erg cm$^{-2}$ s$^{-1}$; for the \xmm\, observation, the parameters are $\Gamma_{\rm X} = 0.91 \pm 0.11$ and $F_{\rm 2-10~keV}=(4.67_{-0.45}^{+0.43}) \times 10^{-13}$ erg cm$^{-2}$ s$^{-1}$. No significant variability in X-ray flux is observed between the two observations; however, there is a notable change in the spectral shape from soft to hard. The same absorbed power-law model is also employed to fit the X-ray spectra reported in previous studies (\citealp{2016ApJ...823...57S, 2019ApJ...871...71S}), and our results are consistent within uncertainties with their findings. It should be noted that there is a noticeable positive residual in the soft X-band region for \xmm\, spectrum. We attempt to improve this by incorporating additional models such as a plasma or power-law component; however, due to limited statistical data, it remains challenging to constrain the model parameters effectively.

\emph{TXS 2352+495}. It was observed once by \xmm\, on 2003 December 25. By fitting the extracted spectrum with an absorbed power-law model, we derive an upper limit of $N_{\rm H}^{\rm int} < 3.6 \times 10^{21}$ cm$^{-2}$ with $\Gamma_{\rm X} = 0.75_{-0.44}^{+0.49}$ and $F_{\rm 2-10~keV}=(8.5_{-3.1}^{+2.2}) \times 10^{-14}$ erg cm$^{-2}$ s$^{-1}$. \cite{2006MNRAS.367..928V} adopted a higher cutoff count rate to exclude time intervals with elevated background activity, thereby obtaining the source spectrum; however, their selected cutoff rate exceeded that utilized in our analysis.

\clearpage

\renewcommand{\thefigure}{A.1}

\begin{figure*}
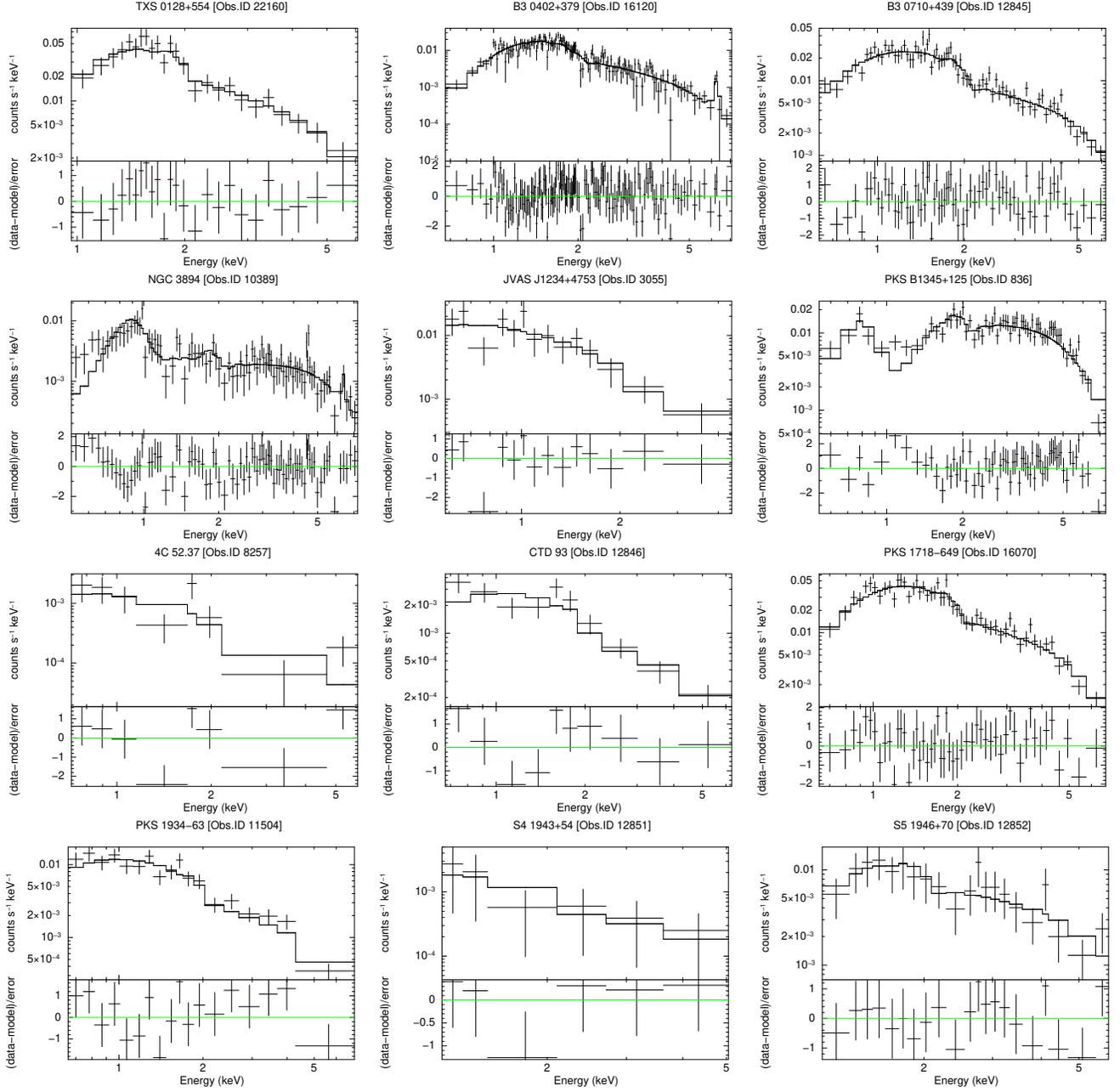

 \centering
   \includegraphics[angle=270,scale=0.21]{chan0128.eps}
   \includegraphics[angle=270,scale=0.21]{chan0402.eps}
   \includegraphics[angle=270,scale=0.21]{chan0710.eps}\\
   \includegraphics[angle=270,scale=0.21]{chan3894.eps}
   \includegraphics[angle=270,scale=0.21]{chan1234.eps}
   \includegraphics[angle=270,scale=0.21]{chan1345.eps}\\
   \includegraphics[angle=270,scale=0.21]{chan52.eps}
   \includegraphics[angle=270,scale=0.21]{chan93.eps}
   \includegraphics[angle=270,scale=0.21]{chan1718.eps}\\
   \includegraphics[angle=270,scale=0.21]{chan1934.eps}
   \includegraphics[angle=270,scale=0.21]{chan1943.eps}
   \includegraphics[angle=270,scale=0.21]{chan1946.eps}
\caption{The \chandra\, spectra for the 12 CSOs listed in Table \ref{tab-Result} are presented along with their respective fitting results. The corresponding residual plots are shown in the lower panels.}
\label{spect-Chandra}
\end{figure*}

\renewcommand{\thefigure}{A.2}

\begin{figure*}
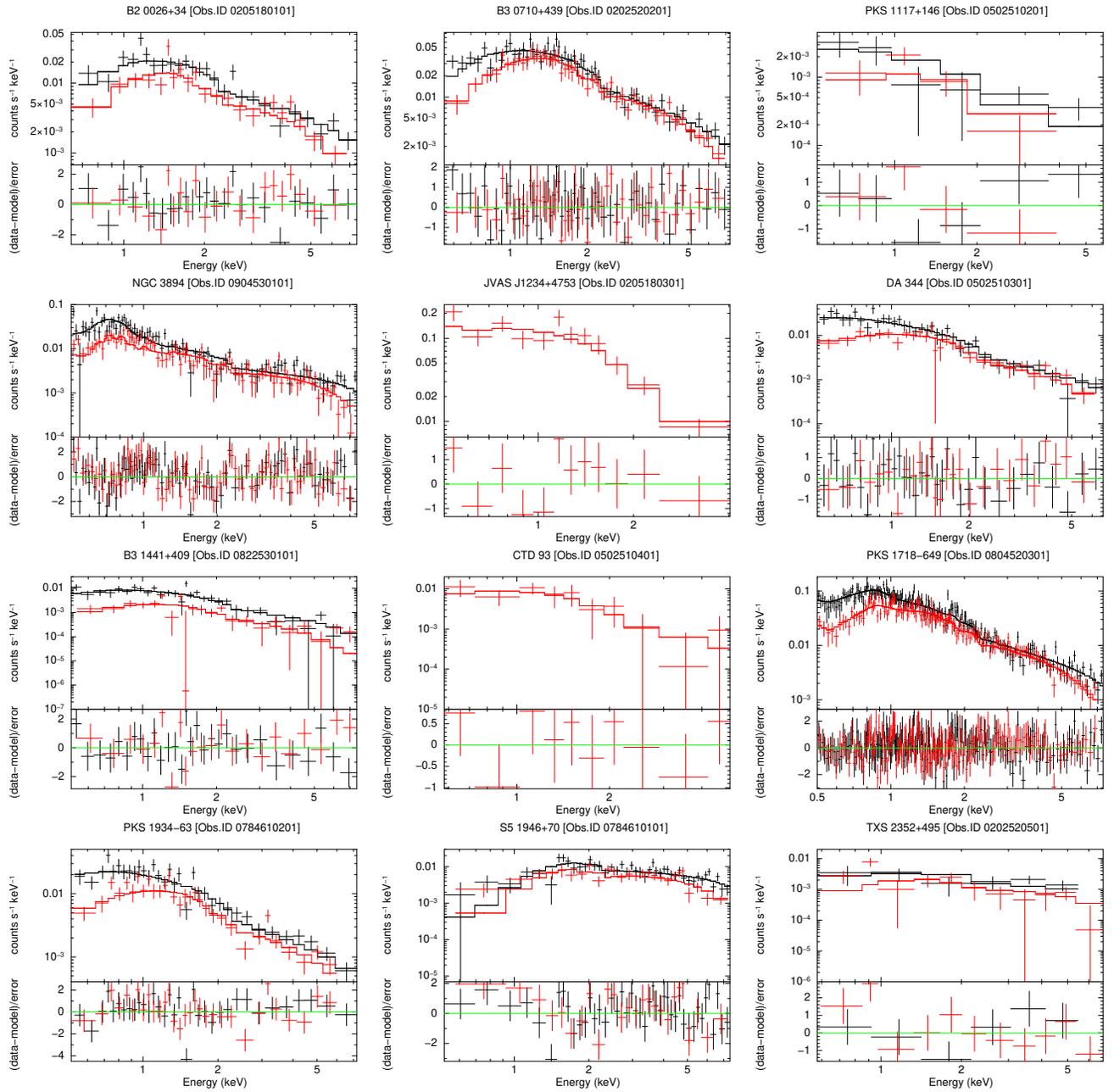

 \centering
   \includegraphics[angle=270,scale=0.21]{XMM0026.eps}
   \includegraphics[angle=270,scale=0.21]{XMM0710.eps}
   \includegraphics[angle=270,scale=0.21]{XMM1117.eps}\\
   \includegraphics[angle=270,scale=0.21]{XMM3894.eps}
   \includegraphics[angle=270,scale=0.21]{XMM1234.eps}
   \includegraphics[angle=270,scale=0.21]{XMM344.eps}\\
   \includegraphics[angle=270,scale=0.21]{XMM1441.eps}
   \includegraphics[angle=270,scale=0.21]{XMM93.eps}
   \includegraphics[angle=270,scale=0.21]{XMM1718.eps}\\
   \includegraphics[angle=270,scale=0.21]{XMM1934.eps}
   \includegraphics[angle=270,scale=0.21]{XMM1946.eps}   
   \includegraphics[angle=270,scale=0.21]{XMM2352.eps}
\caption{The \xmm\, spectra for the 12 CSOs listed in Table \ref{tab-Result} are presented along with their respective fitting results. The corresponding residual plots are shown in the lower panels. The PN spectra are displayed in black, while the MOS spectra are presented in red.}
\label{spect-XMM}
\end{figure*}

\begin{deluxetable}{lccccc}
\tabletypesize{\scriptsize} 
\tablecolumns{6} 
\tablewidth{0pc}
\tablecaption{4 obscured CSOs. Column (1) source name; (2) column density (in 10$^{24}$ cm$^{-2}$); (3) photon spectral index; Column (4) intrinsic 2–10 keV luminosity (in 10$^{43}$ erg s$^{-1}$); Column (5) the fitting results are based on X-ray observation data from these satellites; Column (6) references.}\tablenum{A.1}
\tablehead{\colhead{Source} & \colhead{$N_{\rm H}^{\rm int}$} & \colhead{$\Gamma_{\rm X}$} & \colhead{$L_{\rm 2-10~keV}$} & \colhead{X-ray Satellites} & \colhead{Ref.$^{\blacklozenge}$}\\
\colhead{(1)} & \colhead{(2)} & \colhead{(3)} & \colhead{(4)} & \colhead{(5)} & \colhead{(6)}}
\startdata
OQ 208&0.44$_{-0.02}^{+0.01}$&1.45$_{-0.01}^{+0.11}$&$0.45\pm0.06$&\xmm/\chandra/\emph{NuSTAR}&[1]\\
JVAS J1511+0518&1.3$_{-0.4}^{+0.6}$&1.70$_{-0.05}^{+0.15}$&0.38&\xmm/\emph{NuSTAR}&[2]\\
S4 2021+61&3.7$_{-0.5}^{+0.8}$&1.45$_{-0.05}^{+0.09}$&11&\xmm/\emph{NuSTAR}&[2]\\
NGC 7674&3.4$_{-0.6}^{+0.8}$&2.07$_{-0.11}^{+0.15}$&3--5&\emph{Suzaku}/\emph{Swift}/\emph{NuSTAR}&[3]\\
\hline
\enddata 
\tablenotetext{\blacklozenge}{[1] \cite{2019ApJ...884..166S}; [2] \cite{2023ApJ...948...81S}; [3] \cite{2017MNRAS.467.4606G}.}
\label{tab-4CSOs}
\end{deluxetable}

\end{CJK}
\end{document}